\documentclass[a4paper,10pt]{article}
\usepackage{amsmath,amssymb,mathtools}  
\usepackage{hyperref}
\usepackage{soul}                       
\usepackage[usenames,dvipsnames]{xcolor}
\usepackage[raggedright]{titlesec}
\usepackage{scrextend}
\deffootnote{2em}{0em}{\thefootnotemark\quad}
\usepackage{multicol}
\usepackage[left=0.75in,right=0.75in,top=1.0in,bottom=1.25in]{geometry}

\hypersetup{
    colorlinks,
    linkcolor={red!50!black},
    citecolor={blue},
    urlcolor={blue!80!black}
}

\newcommand{\dd}{\text{d}}
\newcommand{\dt}{\text{d}t}
\newcommand{\dx}{\text{d}x}
\newcommand{\ds}{\text{d}s}

\setstcolor{red}    
\setulcolor{red}    
\allowdisplaybreaks 

\let\oldabstract\abstract
\let\oldendabstract\endabstract
\makeatletter
\renewenvironment{abstract}
{%
               {\list{}{\addtolength{\leftmargin}{4em} 
                        \listparindent 1.5em%
                        \itemindent    \listparindent%
                        \rightmargin   \leftmargin%
                        \parsep        \z@ \@plus\p@}%
                \item\relax}%
               {\endlist}%
\oldabstract}
{\oldendabstract}
\makeatother

\title{\LARGE\textbf{\textsf{Transformation of Primordial Cosmological \\Perturbations Under the General\\Extended Disformal Transformation}}}

\author{\normalsize Allan L. Alinea${}^{(a),(b)} $\footnote{AL Alinea, the corresponding author, is no longer affiliated with Osaka University but part of this work was done during his last years in the said university.}\;\;\;and Takahiro Kubota$ {}^{(b)} $
}

\date{}

\begin{document}

\maketitle
\vspace{-2.25em}
\noindent
\begin{center}
{\small $ {}^{(a)} $Institute of Mathematical Sciences and Physics,\\ University of the Philippines Los Ba\~nos,\\College, Los Ba\~nos, Laguna 4031 Philippines, alalinea@up.edu.ph}
\\[0.5em]
{\small $ {}^{(b)} $Department of Physics, Osaka University\\
Toyonaka, Osaka 560-0043, Japan, 
kubota@celas.osaka-u.ac.jp}
\end{center}

\bigskip
\begin{abstract}
	Primordial cosmological perturbations are the seeds that were cultivated by inflation and the succeeding dynamical processes, eventually leading to the current Universe. In this work, we investigate the behavior of the gauge-invariant scalar and tensor perturbations under the general extended disformal transformation, namely, $g_{\mu\nu} \rightarrow A(X,Y,Z)g_{\mu\nu} + \Phi_\mu\Phi_\nu$, where $X \equiv -\tfrac{1}{2}\phi^{;\mu}\phi_{;\mu}, Y \equiv \phi^{;\mu}X_{;\mu}, Z \equiv X^{;\mu}X_{;\mu} $ and $\Phi_\mu \equiv C\phi_{;\mu} + DX_{;\mu}$, with $C$ and $D$ being a general functional of $(\phi,X,Y,Z)$. We find that the tensor perturbation is invariant under this transformation. On the other hand, the scalar curvature perturbation receives a correction due the conformal term only; it is independent of the disformal term at least up to linear order. Within the framework of the full Horndeski theory, the correction terms turn out to depend linearly on the gauge-invariant comoving density perturbation and the first time-derivative thereof. In the superhorizon limit, all these correction terms vanish, leaving only the original scalar curvature perturbation. In other words, it is invariant under the general extended disformal transformation in the superhorizon limit, in the context of full Horndeski theory. Our work encompasses a chain of research studies on the transformation or invariance of the primordial cosmological perturbations, generalizing their results under our general extended disformal transformation.
\bigskip
\\
\noindent
{\small\textbf{keywords:} \textit{inflation; primordial cosmological perturbations; disformal transformation.}}	
\end{abstract}

\bigskip
\begin{multicols}{2}

\section{Introduction}
\label{sec1}
It is part of our insatiable curiosity about the Universe to peek into the content of the past. In the grand scheme of things, our knowledge of the past allows us to connect it to the present, and when these two are combined, they may enable us to predict what holds for the future. The theory of cosmic inflation \cite{Guth:1980zm, Starobinsky:1980te, Starobinsky:1979ty, Sato:1980yn, Linde:1983gd, Albrecht:1982wi}---a rapid exponential expansion of the early Universe---is nowadays considered an integral part of modern cosmology dealing with the physical ``history'' of the Universe. While inflation could have lasted for a very short amount of time, taking a tiny slice in the spectrum of the past, the theory finds its relevance in solving the relatively long standing horizon and flatness problems that were then plaguing the standard Big Bang cosmology \cite{Dodelson:2003ft}. Furthermore, inflation takes us to the fundamental level of our main interest in this work namely, primordial cosmological perturbations---the seeds that gave birth to what we nowadays observe as galaxies and clusters of galaxies. 

Although cosmic inflation may appear as a rather simple phenomenon of a rapidly expanding spacetime, considering how it could be brought about and end, and the possible differences in the dynamics of quantum field(s) involved, there is a plethora of models describing it. For single-field inflation alone, the plurality may include the simplest canonical single field inflation \cite{Maldacena:2002vr, Mukhanov:2005sc}, the kinetically driven inflation involving a ``pressure'' functional \cite{ArmendarizPicon:1999rj,Garriga:1999vw}, and inflation models with non-minimally coupled term(s) \cite{Fakir:1990eg, Futamase:1987ua}, amongst others. In spite of the multitude of models however, all of them have to take into account the primordial cosmological perturbations \cite{Weinberg:2008zzc, Gorbunov:2011zz} and connect them to something that can be observed or measured; e.g., power spectrum, tensor-to-scalar ratio, etc. The final arbiter of an ensemble of competing theories is always a good set of experiments.

Having said this, we have plenty of reasons to also look into the mathematical aspects of primordial cosmological perturbations as one goes from one theory to another, perhaps with greater level of generality. One may see for instance, that these perturbations are invariant under conformal transformation in dealing with a non-minimally coupled inflation models; see for instance, Ref. \cite{Makino:1991sg,Chiba:2008,Gong:2011} . Using this knowledge may allow one to perform calculations in the more general model with ease and simplicity as in the derivation of non-gaussianity \cite{Qiu:2010dk,Kubota:2011re}, adiabatic regularization of power spectrum \cite{Alinea:2016qlf,Alinea:2017ncx}, or the mere calculation of power spectrum. Furthermore, knowledge of invariance or transformation properties of primordial cosmological perturbations may facilitate in relating one model of inflation to another. From a generalist perspective, it is also interesting to understand how these perturbations may  vary (or remain invariant) under a general transformation within the framework of a general inflation model; thus, allowing us to make general insight(s) or conclusion(s) on a possibly wide array of inflation theories contained within the general framework under consideration.

The (covariantized) Galileon inflation \cite{Deffayet:2009mn,Deffayet:2011gz,Kase:2014cwa} is the most general single-field model of inflation with a second-order corresponding set of equations of motion consistent with the requirement to evade the Ostrogradsky instability \cite{Woodard:2015zca}. The corresponding action is actually a ``rediscovery''\cite{Kobayashi:2011nu,Kobayashi:2019hrl} of the Horndeski action \cite{Horndeski:1974wa, Charmousis:2011bf} established about four decades earlier. In Ref. \cite{Bettoni:2013diz}, the authors showed that the Horndeski action is form-invariant under the transformation
 \cite{Bekenstein:1992pj} of the metric $ g_{\mu \nu } $, that we call here \textit{special disformal transformation}, given by
\begin{align}
	\label{DTbasic}
	g_{\mu \nu }
	\rightarrow
	\widehat g_{\mu \nu }
	=
	A(\phi )g_{\mu \nu } + B(\phi )\phi _{;\mu }\phi _{;\nu },
\end{align}
where $ \phi $ is a scalar inflaton field, $(A,B)$ are functionals of $ \phi $ and subject to some (physical) constraints (e.g., invertibility of the metric and causality), and the semicolon denotes covariant derivative. Interestingly, viewing the Horndeski theory as an inflation theory, it was shown that the gauge-invariant scalar and tensor  perturbations are invariant under the special case of disformal transformation given above \cite{Minamitsuji:2014waa}. 

This work aims to look into the transformation (or invariance) of the gauge-invariant primordial cosmological perturbations \cite{Bardeen:1980kt}---in particular, the scalar curvature and tensor perturbations---under the general extended disformal transformation encompassing (\ref{DTbasic}), given by
\begin{align}
	\label{genextdt}
	g_{\mu \nu }
	\rightarrow
	\widehat g_{\mu \nu }
	=
	A(\phi, X,Y,Z) g_{\mu \nu }
	+
	\Phi _{\mu } \Phi _{\nu },
\end{align}
where 
\begin{align}
	\label{defXYZ}
	X
	&\equiv 
	-\tfrac{1}{2}g^{\mu \nu }\phi _{;\mu }\phi _{;\nu },
	\nonumber
	\\[0.5em]
	Y
	&\equiv
	g^{\mu \nu }\phi _{;\mu }X_{;\nu },
	\nonumber
	\\[0.5em]
	Z
	&\equiv
	g^{\mu \nu }X_{;\mu}X_{;\nu },
\end{align}
with\footnote{Note that terms $ Y $ and $ Z $ are generated in the transformation of the scalar curvature, Ricci tensor, and $ (\square \phi - \phi ^{;\mu \nu }\phi _{;\mu \nu })$, amongst others, under the special disformal transformation; however, they are canceled by other terms in the Horndeski action.}
\begin{align}
	\Phi _\mu 
	\equiv
	C(\phi, X,Y,Z)\phi _{;\mu }
	+
	D(\phi, X,Y,Z)X_{;\mu }.
\end{align}

The extended disformal transformation (\ref{genextdt}) is an umbrella transformation encompassing several important special cases considered in the literature; see the following paragraph. As already mentioned, we would like to know under our general framework, the behavior of the gauge-invariant primordial cosmological perturbations. These gauge-invariant quantities, namely, the scalar and tensor perturbations correspond to CMB and primordial gravitational wave power spectra, respectively, which are physical observables that serve as our window into the early universe. 

This study may be seen as a part of a ``series'' of articles (see \textbf{Table \ref{table01}}) with different groups of authors investigating the transformation properties of the primordial cosmological perturbations under more and more general disformal transformation. After the publication of Ref. \cite{Minamitsuji:2014waa} wherein $ A $ and $ B $ are both taken as functionals of $ \phi  $ alone, Ref. \cite{Tsujikawa:2014uza} showed that when $ B $ is a functional of both $ (\phi ,X) $ while keeping $ A = A(\phi ) $, the scalar and tensor perturbations remain invariant at least at linear order.\footnote{The result in Ref. \cite{Domenech:2015hka} wherein the cosmological perturbations are found to be invariant under pure disformal transformation namely, $ g_{\mu \nu } \rightarrow g_{\mu \nu } + B(\phi ,X)\phi _{;\mu }\phi _{;\nu }$, may be considered as intermediate between those in Refs. \cite{Minamitsuji:2014waa,Tsujikawa:2014uza}} Following this is Ref. \cite{Motohashi:2015pra} (see also Ref. \cite{Bordin:2017hal}) which further generalised the result in Ref. \cite{Minamitsuji:2014waa} by considering a general disformal transformation where $ A = A(\phi ,X) $ and $ B = B(\phi ,X) $. This time the curvature perturbation is found to be no longer invariant. However, the correction term in the expression for the disformally transformed perturbation turns out to be proportional to some slow-roll parameter. At superhorizon scales, this correction term vanishes. Since physical observables such as power spectrum and tensor-to-scalar ratio involve superhorizon modes, they remain invariant under the general disformal transformation in Ref. \cite{Motohashi:2015pra} where both $ A $ and $ B $ depend on $ (\phi ,X)$. 

\begin{table*}[t]
\centering
\begin{tabular}{ll}
\textbf{References} &\qquad \textbf{Disformal Transformation $ (\widehat {\boldsymbol g}_{\boldsymbol\mu \boldsymbol\nu }) $}\\[0.25em] \hline \\[-0.25em]
Ref. \cite{Minamitsuji:2014waa} 
& 
$ 
\qquad\quad\;\; A(\phi ) g_{\mu \nu } + \; B(\phi )\phi _{;\mu }\phi _{;\nu }$\\[0.5em] 
Ref. \cite{Domenech:2015hka} 
& 
$ 
\qquad\qquad\quad\;\;\, g_{\mu \nu } +\; B(\phi,X )\phi _{;\mu }\phi _{;\nu }$\\[0.5em] 
Ref. \cite{Tsujikawa:2014uza} 
& 
$ 
\qquad \quad\;\; A(\phi )g_{\mu \nu } +\; B(\phi,X )\phi _{;\mu }\phi _{;\nu }$\\[0.5em] 
Ref. \cite{Motohashi:2015pra} 
& 
$ 
\qquad \;A(\phi,X )g_{\mu \nu } +\; B(\phi,X )\phi _{;\mu }\phi _{;\nu }$\\[0.5em] 
current work 
& 
$ 
A(\phi,X,Y,Z)g_{\mu \nu } 
+ 
(C\phi _{;\mu } + DX_{;\mu })
(C\phi _{;\nu } + DX_{;\nu })$\\ [0.5em]\hline
\end{tabular}
\caption{Series of researches leading to the current work on disformal transformation and primordial cosmological perturbations.}
\label{table01}
\end{table*}

Along the line of these developments we would like to show  the following two facts in the present paper:
\begin{description}
\item{(a)}
Under the transformation (\ref{genextdt}), the gauge invariant curvature perturbation receives corrections, {to linear order,} only due to the conformal term $A$. In other words, the corrections do not depend on the disformal part of  (\ref{genextdt}) in spite of its rather involved form.
\item{(b)}
The corrections in the curvature perturbation can be expressed in terms of the comoving density perturbation and its time-derivative which disappear on the superhorizon scale.
\end{description}

These two findings, (a) and (b), give us a strong impetus to look  for a more general class of scalar-tensor theories with the aid of (\ref{genextdt}) than those considered in literatures for future studies. The form of the general extended disformal transformation given 6by (\ref{genextdt}) can be somewhat badly intriguing. For one thing, this transformation will certainly ``push'' the Horndeski action outside the set of possible Horndeski forms (with second order equations of motion). Nonetheless, it is apparent in Ref. \cite{Bettoni:2013diz} that even the slightest addition of dependency of either $ A $ or $ B $ on $ X $ will make the transformed Horndeski action no longer of the original Horndeski form. And we find somewhat of an excuse in Refs. \cite{Tsujikawa:2014uza,Motohashi:2015pra} wherein the $X$ dependence is considered in the investigation of the disformal transformation of primordial cosmological perturbations. We now know that the addition of $ X $-dependence in the transformation of the Horndeski action is not necessarily a bad thing even if the equations of motion can exceed second order, as long as there are suitable constraints, as in the GLPV theory \cite{Gleyzes:2014qga,Gleyzes:2014dya} (see also Refs. \cite{Zumalacarregui:2013pma,Langlois:2015cwa,Crisostomi:2016czh,BenAchour:2016fzp}), that can be imposed to prevent the existence of ghosts. 

For the case at hand, the existence of ghost/Ostrogradsky instability 
in the transformed Horndeski action is apparent in the absence of suitable constraints or conditions and it is beyond the scope of the present paper to get rid of the instabilities; although see the appendix for some insights. The main focus is on our modest objective to look into the transformation of the primordial cosmological perturbations---the gauge-invariant scalar and tensor perturbations in particular---under the general disformal transformation stated above, as discussed in  Sec. \ref{sec2}. We argue in  Sec. \ref{sec3} that {to linear order,} only the conformal part $A$ affects the curvature perturbation. {Furthermore,} the corrections are expressed in terms of the comoving density perturbation and its time derivative. In Sec. \ref{limigextdt}, we present a short discussion on the invertibility of the general extended disformal transformation. Our conclusions and  remarks {are stated} in Sect. \ref{conclude}. This is then followed by a rather lengthy appendix where the transformation formulas of some important terms in the Horndeski action are collected. 

\bigskip
\section{Transformation of the Gauge-Invariant Scalar and Tensor Perturbations}
\label{sec2}

We start the calculation\footnote{Our calculation here follows the same logic by Motohashi and White presented in Ref. \cite{Motohashi:2015pra}.
}   
of the possible transformation of the primordial cosmological perturbations, under the general extended disformal transformation given by (\ref{genextdt}), by considering a perturbed metric (involving SVT-decomposed perturbations) about the Friedmann-Lema\^itre-Robertson-Walker (FLRW) background spacetime, given by\footnote{We are using the metric signature $ (-,+,+,+) $.} 
\begin{align}
	\label{flrwperturbds}
	\ds^2
	&=
	-(1 + 2\varphi )\dt^2
	+
	2a(\alpha _{,i} + \beta _{i})\dt\,\dx^i
	\\[0.5em]
	&\quad
	+\,
	a^2\big[
		(1 - 2\psi )\delta _{ij} + \gamma _{ij} + 2E_{,ij} + 2F_{(i,j)}
	\big]
	\dx^i \dx^j,
	\nonumber
\end{align}
where $ a = a(t) $ is the scale factor, $ (\varphi , \alpha , \psi ,E) $ are scalar perturbations, and the vector $ (\beta _i, F_i)  $ and tensor perturbations ($ \gamma _{ij} $) satisfy the following conditions:
\begin{align}
	\partial ^i\beta _i = \partial^i F_i = 0
	\quad
	\text{and}
	\quad
	\gamma ^i{}_i = \partial ^i\gamma _{ij} = 0. 
\end{align}
The comma in the expression for the perturbed metric above signifies partial differentiation.

Under the general extended disformal transformation given by (\ref{genextdt}), the transformed metric that we write here with a ``hat'', becomes
\begin{align}
	\dd\widehat s^{\,2}
	&=
	\widehat g_{\mu \nu }\dx^\mu \dx^\nu ,
	\nonumber
	\\[0.5em]
	\dd\widehat s^{\,2}
	&=
	(Ag_{00} + \Phi _0^2)\dt^2
	+
	2(Ag_{0i} + \Phi _0\Phi _i)\dt\dx^i
	\nonumber
	\\[0.5em]
	&\quad	
	+\,
	(Ag_{ij} + \Phi _i\Phi _j)\dx^i\dx^j.
\end{align}
The field $ \phi  $ in the functional $ A $ can be expressed to linear order as a sum of its background value (denoted with a bar) and a perturbation about it, namely, $ \phi (t,\vec x) = \bar \phi (t) + \delta \phi (t,\vec x) $. By virtue of the definition for $X$, this has as a consequence, 
\begin{align}
	\label{expX}
	X 
	= 
	\bar X + \delta X 
	= 
	\tfrac{1}{2}{\dot{\bar \phi }}^2 
	- 
	\big(
		{\dot{\bar \phi }}^2 \varphi   
		- 
		\dot {\bar \phi }\,\delta \dot \phi 
	\big).
\end{align}
Furthermore, $\Phi _0 = \bar \Phi _0 + \delta \Phi _0$, with $ \delta \Phi _0 $ being a first-order perturbation, while $ \Phi _i $ is already first order. The quantity $ A $ in the disformal transformation can also be written in the same way as $ A = \bar A + \delta A $. Consequently, the transformed metric becomes
\begin{align}
	\label{flrwperturbdsh}
	\dd\widehat s^{\,2}
	&=
	-(
		\bar A + \delta A 
		+ 
		2\bar A\varphi - \bar \Phi _0^2 
		- 
		2\bar \Phi _0\,\delta \Phi _0
	)\dt^2
	\nonumber
	\\[0.5em]
	&\quad 
	+\,
	2a{\bar A}^{\frac{1}{2}}\big[
		{\bar A}^{\frac{1}{2}}(\alpha _{,i} + \beta _i)
		+
		\bar \Phi _0\Phi _i/(a{\bar A}^{\frac{1}{2}})
	\big] \dt\dx^i
	\nonumber
	\\[0.5em]
	&\quad
	+
	a^2\bar A\big[
		(1 - 2\psi +\delta A/\bar A)\delta _{ij}
		+
		\gamma _{ij}
		\nonumber
		\\[0.5em]
		&\qquad\qquad\quad
		+\,
		2E_{,ij}
		+
		2F_{(i,j)}
	\big]\dx^i\dx^j.
\end{align}

\medskip
If we switch off the perturbations, the hatted metric becomes $\dd\widehat s^{\,2} = -(\bar A - \bar \Phi _0^2)\dt^2 + a^2\bar A\,\delta _{ij}\,\dx^i\dx^j$, which, following the form of the background FLRW metric, is suggestive of coordinate time and scale factor scaling. Following the idea for pure conformal transformation as in Ref. \cite{Kubota:2011re}, we then define the hatted time coordinate and scale factor as
\begin{align}
	\label{hatTna}
	\dd\widehat t^{\,2}
	\equiv
	(\bar A - {\bar \Phi _0}^2)\dt^2
	\quad
	\text{and}
	\quad
	\widehat a
	\equiv
	a{\bar A}^{\frac{1}{2}},  
\end{align}
With these definitions, together with the hatted perturbation variables, we may alternatively write the transformed metric as
\begin{align}
	\dd\widehat s^{\,2}
	&=
	-(1 + 2\widehat \varphi )\dd\widehat t^{\,2}
	+
	2\widehat a(\widehat \alpha _{,i} + \widehat \beta _i)
	\dd\widehat t\,\dx^i
	\\[0.5em]
	&\quad
	+\,
	\widehat a^2\big[
		(1 - 2\widehat \psi )\delta _{ij}
		+
		\widehat \gamma _{ij}
		+
		2\widehat E_{,ij}
		+
		2\widehat F_{(i,j)}
	\big]\dx^i\dx^j.
	\nonumber
\end{align}
Comparing the two equations above for the transformed metric will allow us to find the relationships between the transformed perturbation variables and the original ones.
\medskip

For the 00-component we find
\begin{align}
	\widehat \varphi 
	=
	\bigg(
		\varphi 
		+
		\frac{\delta A}{2\bar A}
		-
		\frac{\bar \Phi _0}{\bar A}
		\delta \Phi _0  
	\bigg)\bigg(
		1 - \frac{\bar \Phi _0^2}{\bar A} 
	\bigg)^{-1}.
\end{align}
For the $0i$-components we have
\begin{align}
	2\widehat a(\widehat \alpha _{,i} + \widehat \beta _i)
	&=
	\frac{2\widehat a\big[
		{\bar A}^{\frac{1}{2}}(\alpha _{,i} 
		+ 
		\beta _i) + \bar \Phi _0\Phi _i/(a\bar A^{\frac{1}{2}})
	\big]}{(\bar A - \bar \Phi _0^2)^{1/2} } 
\end{align}
Noting that $ \Phi _i $ can be expressed to linear order as $ \Phi _i = \bar C\phi _{,i} + \bar D\,\delta X_{,i} $, where $ \bar C $ and $ \bar D $ are the background values of $ C $ and $ D $ respectively, we can decompose the above equation for $ \widehat \alpha  $ and $ \widehat \beta_i  $ as
\begin{align}
	\widehat \alpha 
	&=
	\bigg[
		\alpha 
		+
		\frac{\bar \Phi _0}{a\bar A}\big(
			\bar C\,\delta \phi 
			+
			\bar D\,\delta X
		\big) 
	\bigg]\bigg(
		1 - \frac{\bar \Phi _0}{\bar A} 
	\bigg)^{-\frac{1}{2} },
	\nonumber
	\\[0.5em]
	\widehat \beta _i
	&=
	\beta _i\bigg(
		1 - \frac{\bar \Phi _0}{\bar A} 
	\bigg)^{-\frac{1}{2} }.
\end{align}
Lastly, for the $ ij $-components, comparison of the (spatial parts of the) two hatted metrics yield
\begin{align}
	\label{psihatetc}
	\widehat \psi 
	&=
	\psi - \frac{\delta A}{2\bar A},
	\nonumber
	\\[0.5em]
	\widehat \gamma _{ij}
	&=
 	\gamma _{ij},
 	\nonumber
 	\\[0.5em]
 	\widehat E
 	&=
 	E,
 	\nonumber
 	\\[0.5em]
 	\widehat F
 	&=
	F.
\end{align}
We thus, find that of the two perturbations we are mainly interested in, namely, the gauge-invariant scalar and tensor perturbations, the latter is invariant under the general extended disformal transformation, at least up to linear order. Needless to say, this translates to invariance of the tensor power spectrum, graviton non-gaussianity \cite{Maldacena:2011nz}, etc., under the general extended disformal transformation given by (\ref{genextdt}).
\medskip

We now move on to the gauge-invariant scalar curvature perturbation. It is defined as
\begin{align}
	\label{defRc}
	{\mathcal R}_c
	&\equiv
	-\psi  - \frac{H}{\dot{\bar \phi }} \delta \phi,
\end{align}
where $ H $ is the Hubble parameter given by $ H \equiv \dot a/a, $ with the dot indicating ordinary time derivative. Corresponding to this is the disformally transformed (hatted) curvature perturbation given by
\begin{align}
	\widehat {\mathcal R}_c
	&=
	-\widehat \psi 
	-
	\frac{\widehat H}{\dd\bar \phi /\dd\widehat t} \,
	\delta \phi,
\end{align}
where $ \widehat H = (\dd\widehat a/\dd\widehat t)/\widehat a $. Observe that $ \widehat H $ expressed in relation to $ H $, involves the (background) disformal term $ \bar \Phi _0 $, by virtue of (\ref{hatTna}) relating $ \dd\widehat t $ and $ \dt $.
\begin{align}
	\label{Hhat}
	\widehat H
	&=
	\bigg(
		H + \frac{\dot{\bar A}}{2\bar A} 
	\bigg)\big(
		\bar A - \bar \Phi _0^2
	\big)^{-\frac{1}{2} }.
\end{align}
On the other hand, $ \widehat \psi  $ by virtue of (\ref{psihatetc}), carries the dependence on the conformal term.  However, for $\widehat H $, because $ \widehat {\mathcal R}_c $ involves the ratio of $ \widehat H $ and $ \dd\bar \phi /\dd\widehat t $, the second factor on the right hand side of (\ref{Hhat}) cancels. Consequently,
\begin{align}
	\label{hattedR}
	\widehat {\mathcal R}_c
	&=
	\mathcal R_c
	+
	\frac{\delta A}{2\bar A} 
	-
	\frac{\dot{\bar A}}{2\bar A}
	\frac{\delta \phi }{\dot{\bar \phi }}.   
\end{align}
The gauge-invariant curvature perturbation $ \widehat {\mathcal R} _c$ expressed in relation to $ \mathcal R_c $, does \textit{not} explicitly depend on the disformal term in the general extended disformal transformation (\ref{genextdt}); the added term to $ \mathcal R_c $ depends only on the conformal term $A$, field $ \phi  $, and its variation. The absence of the disformal term appears as a sort of general characteristic of the disformally transformed gauge-invariant curvature perturbation (at least up to linear order) expressed in terms of the original perturbation. With the assumption of FLRW background spacetime and the inflaton decomposition $ \phi (t,\vec x) = \bar \phi (t) + \delta \phi (t,\vec x) $, we can make our general extended disformal transformation even more general by modifying the disformal term, say, including dependence of $ C $ and $ D $ on $ X^{;\mu \nu } X_{;\mu \nu } $; the consequence will nonetheless not affect the transformation of $ \widehat {\mathcal R}_c $ at least at the linear level. The correction term, as what we find, is mainly anchored on the conformal part of the general extended disformal transformation.
\medskip

Ref. \cite{Domenech:2015hka} emphasized that pure disformal transformation, one wherein $ A = 1 $, may be interpreted as a sort of time rescaling. This interpretation can be carried over to our more general treatment with a slight modification; that is, accompanying time rescaling is the rescaling of the scale factor by the conformal factor (see (\ref{hatTna})). This is apparent at the background level wherein the transformed metric may be written as
\begin{align}
	\dd\widehat s^{\,2}
	=
	\dd\widehat t^{\,2}
	+
	\widehat a^{\,2}\,\delta _{ij}\dx^i\dx^j,
\end{align}
following the form of the original background FLRW metric. Now, looking at the definition of the gauge-invariant scalar perturbation given by (\ref{defRc}), we find that the second term is time-rescaling invariant at the first order. On the other hand, the first term, being a perturbation in the \textit{spatial} part of the metric, receives a correction due to the conformal factor; see (\ref{flrwperturbdsh}). The overall effect then of disformal transformation on $ \mathcal R_c $, seen as a disformal time and conformal scale factor rescalings, is a transformed perturbation with a correction that depends only on the conformal factor, at the linear level.

\bigskip
\section{Correction Terms in the Gauge-Invariant Curvature Perturbation}
\label{sec3}
Looking at the equation for $ \widehat {\mathcal R}_c $ given by (\ref{hattedR}), we see that the only term that requires our attention for simplification is the second term on the right hand side involving the variation of conformal factor $ \delta A $. Since $ A $ is a functional of $(\phi ,X,Y,Z)$, this variation divided by $\bar A$ can be ``spelled out'' as
\begin{align}
	\label{deltaA}
	\frac{\delta  A}{\bar A} 
	&=
	\frac{\bar A_{,\bar \phi }}{\bar A} \,\delta \phi 
	+
	\frac{\bar A_{,\bar X}}{\bar A} \,\delta X
	+
	\frac{\bar A_{,\bar Y}}{\bar A} \,\delta Y
	+
	\frac{\bar A_{,\bar Z}}{\bar A} \, \delta Z.
\end{align}
For the first term on the right hand side, from the equation for the ordinary derivative of $\bar A$ with respect to time, we can write the equation for $ \bar A_{,\bar \phi }/\bar A $ as
\begin{align}
	\frac{\bar A_{,\bar \phi }}{\bar A} 
	&=
	\frac{\dot{\bar A}}{\dot {\bar \phi }\bar A} 
	-
	\frac{\bar A_{,\bar X}}{\bar A} 
	\frac{\dot{\bar X}}{\dot {\bar \phi }} 
	-
	\frac{\bar A_{,\bar Y}}{\bar A} 
	\frac{\dot{\bar Y}}{\dot {\bar \phi }} 
	-
	\frac{\bar A_{,\bar Z}}{\bar A} 
	\frac{\dot{\bar Z}}{\dot {\bar \phi }}.
\end{align}
The variations of $ X, Y, $ and $ Z $ in the last three terms in (\ref{deltaA}) can be computed following their definitions given by (\ref{defXYZ}). With the equation for $ X $ already given by (\ref{expX}), we have for $ Y $ and $ Z $ to linear order,
\begin{align}
	\label{YdelYZdelZ}
	Y
	&=
	\bar Y
	+
	\delta Y,
	\\[0.5em]
	&=
	-{\dot{\bar \phi }}^2\ddot{\bar \phi }
	+
	\big(
		4{\dot{\bar \phi }}^2\ddot {\bar \phi }\varphi 	
		+
		{\dot{\bar \phi }}^3\dot \varphi 	
		-
		2\dot {\bar \phi }\ddot{\bar \phi }\,\delta \dot \phi 
		-
		{\dot{\bar \phi }}^2\,\delta \ddot\phi 	
	\big),
	\nonumber
	\\[0.5em]
	Z
	&=
	\bar Z + \delta Z,
	\nonumber
	\\[0.5em]
	&=
	-{\dot{\bar \phi }}^2{\ddot{\bar \phi }}^2
	+
	\big(
		6{\dot{\bar \phi }}^2{\ddot{\bar \phi }}^2\varphi 
		+
		2{\dot{\bar \phi }}^3\ddot{\bar \phi }\dot \varphi 
		-
		2{\dot{\bar \phi }}{\ddot{\bar \phi }}^2\,\delta \dot \phi 
		-
		2{\dot{\bar \phi }}^2\ddot{\bar \phi }\,\delta \ddot\phi 
	\big).
	\nonumber
\end{align}

Using the background values of $ X, Y,\, $ and $ Z $ to compute for $ \dot{\bar X}, \dot{\bar Y},$ and $ \dot {\bar Z} $, in the equation for $ \bar A_{,\bar \phi }/\bar A $ above, we find 
\begin{align}
	\frac{\bar A_{,\bar \phi }}{\bar A}
	&=
	\frac{\dot{\bar A}}{\dot {\bar \phi }\bar A} 
	-
	\frac{\bar A_{,\bar X}}{\bar A} 
	\ddot{\bar \phi }
	+
	\frac{\bar A_{,\bar Y}}{\bar A} 
	\big(
		2{\ddot{\bar \phi }}^2 + \dot{\bar \phi }\,\dddot{\bar \phi }
	\big)
	\nonumber
	\\[0.5em]
	&\quad
	+\,
	2\frac{\bar A_{,\bar Z}}{\bar A} \ddot{\bar \phi }
	\big(
		{\ddot{\bar \phi }}^2
		+
		\dot{\bar \phi }\,\dddot{\bar \phi }
	\big).
\end{align}
Performing substitution from this equation, and (\ref{expX}) and (\ref{YdelYZdelZ}) for $ \delta X $ and $ (\delta Y, \delta Z) $ respectively, in (\ref{deltaA}) for $ \delta A $, we find upon using the result in the equation for $ \widehat {\mathcal R}_c $ given by (\ref{hattedR})
\begin{align}
	\label{Rccola}
	\widehat {\mathcal R}_c
	&=
	\mathcal R_c
	-
	\frac{\bar A_{,\bar X}}{2\bar A} \big(
		{\dot{\bar \phi }}^2\varphi 
		+
		\ddot{\bar \phi }\,\delta \phi 
		-
		\dot {\bar \phi }\,\delta \dot \phi 
	\big)
	\nonumber
	\\[0.5em]
	&\quad\quad\;
	+\,
	\frac{\bar A_{,\bar Y}}{2\bar A}\big[
		{\dot{\bar \phi }}^2\big(
			4\ddot{\bar \phi }\varphi 
			+
			\dot {\bar \phi} \dot \varphi 
		\big)
		+
		\big(
			2{\ddot{\bar \phi }}^2 
			+ 
			\dot{\bar \phi }\,\dddot{\bar \phi }		
		\big)\,\delta \phi
		\nonumber
		\\[0.35em]
		&\qquad\qquad\qquad
		-\,
		2\dot{\bar \phi} \ddot{\bar \phi }\,\delta \dot \phi 		
		-
		{\dot{\bar \phi }}^2\,\delta \ddot \phi 
	\big]
	\nonumber
	\\[0.5em]
	&\quad\quad\;
	+\,
	\frac{\bar A_{,\bar Z}}{\bar A}\ddot {\bar \phi }\big[
		{\dot{\bar \phi }}^2\big(
			3\ddot{\bar \phi }\varphi + \dot{\bar \phi} \dot \varphi 
		\big)
		+
		\big(
			{\ddot{\bar \phi }}^2
			+
			\dot {\bar \phi }\,\dddot{\overline \phi }
		\big)\,\delta \phi 
		\nonumber
		\\[0.35em]
		&\qquad\qquad\qquad
		-\,
		\dot{\bar \phi }\big(
			\ddot{\bar \phi }\,\delta \dot \phi 
			+
			\dot{\bar \phi  }\,\delta \ddot \phi
		\big)
	\big].
\end{align}

In the limit where the conformal factor does not depend on both $ (Y,Z) $, the general extended disformal transformation given by (\ref{genextdt}) reduces to 
\begin{align}
	\label{specExtDT}
	g_{\mu \nu }
	\rightarrow
	\widehat g_{\mu \nu }
	=
	A(\phi ,X)g_{\mu \nu }
	+
	\Phi _\mu \Phi _\nu ,
	\quad
	[\Phi _\mu = C(\phi ,X)\phi _{;\mu } + D(\phi ,X)X_{;\nu }],
\end{align}
which is still an effective ``superset'' of the general disformal transformation in Ref. \cite{Motohashi:2015pra} where $ D = 0 $. As mentioned above however, the presence of $ D $ does not affect the transformation of $ \mathcal R_c $ (and the tensor perturbation $ \gamma _{ij} $) at least up to linear order. With the special case given by (\ref{specExtDT}), the equation for the curvature transformation reduces to
\begin{align}
	\label{motowhite}
	\widehat {\mathcal R}_c
	&=
	\mathcal R_c
	-
	\frac{\bar A_{,\bar X}}{2\bar A} \big(
		{\dot{\bar \phi }}^2\varphi 
		+
		\ddot{\bar \phi }\,\delta \phi 
		-
		\dot {\bar \phi }\,\delta \dot \phi
	\big),
\end{align}
which is the same result in Ref. \cite{Motohashi:2015pra} wherein $ D = 0 $. Furthermore, it is apparent that when the conformal factor has no dependence on $ X $, the curvature perturbation is disformally invariant within the special contexts in Refs. \cite{Minamitsuji:2014waa,Tsujikawa:2014uza} which are effectively special cases of Ref. \cite{Motohashi:2015pra}.

In their study of adiabatic and entropy perturbations within the framework of multi-field scalar field inflation, the authors of Ref. \cite{Gordon:2000hv} with respect to the original formulation of gauge-invariant variables in Ref. \cite{Bardeen:1980kt} , identified the negative of the expression inside the pair of parentheses in (\ref{motowhite}) as a gauge-invariant comoving density perturbation, namely,
\begin{align}
	\label{eminem}
	\epsilon _m 
	&=
	\dot {\bar \phi }\,\delta \dot \phi
	-
	{\dot{\bar \phi }}^2\varphi 
	-
	\ddot{\bar \phi }\,\delta \phi.
\end{align}
Within the inflation framework they investigated, they showed that based on energy and momentum constraints, $ \epsilon _m \sim k^2/a^2 $, where $ k $ is the comoving wavenumber. It follows that in the same framework, for superhorizon modes, $ \epsilon _m \sim 0$. This is consistent with later work in Ref. \cite{Weinberg:2003sw} tackling adiabatic modes in canonical single-field inflation and multi-field inflation. Within the framework of full Horndeski theory in the context of cosmic inflation, Ref. \cite{Motohashi:2015pra} showed that $ \epsilon _m \sim k^2/a^2H^2$, as in the canonical inflation (see Eq. (2.29) in the mentioned reference); consequently, $ \epsilon _m $ vanishes in the superhorizon limit. Alternatively, the authors found that based on momentum constraint, $ \epsilon _m \sim \dot {\mathcal R}_c $, indicating that an alternative sufficient condition for $ \epsilon _m $ to vanish, and hence, for the relation $ \widehat {\mathcal R}_c = \mathcal R_c$ to hold, is for $ \mathcal R_c $ to remain constant. This is consistent with the first condition since $ \mathcal R_c $ approaches a constant value in the superhorizon limit.

For the case at hand, the conformal factor has $ Y $- and $ Z $- dependencies in addition to $ (\phi ,X) $ and needless to say, we have a more general and more complicated relationship given by (\ref{Rccola}). With the explanations in the preceding paragraphs, the first correction term is no longer a problem as it is guaranteed to vanish in the superhorizon limit at least in the full Horndeski theory. For the second and third correction terms, we are fortunate for the last two terms in (\ref{hattedR}) involving $ \delta A $ to ``conspire'' with the definitions of their dependencies on $ (X,Y,Z) $ given by (\ref{defXYZ}). As it turns out, with reference to the definition of $ \epsilon _m $ given by (\ref{eminem}), we can rewrite the equation for $ \widehat {\mathcal R}_c $ given by (\ref{Rccola}) as
\end{multicols}

\vspace{-2.0em}
\begin{align}
	\label{Rcsimp}
	\boxed{
	\widehat {\mathcal R}_c
	=
	\mathcal R_c
	+
	\frac{\bar A_{,\bar X}}{2\bar A}
	\epsilon _m
	-
	\frac{\bar A_{,\bar Y}}{2\bar A}\big(
		2\ddot{\bar \phi }\epsilon _m
		+
		\dot{\bar \phi }\dot\epsilon _m
	\big)
	-
	\frac{\bar A_{,\bar Z}}{\bar A}\ddot {\bar \phi }\big(
		\ddot{\bar \phi }\epsilon _m + \dot{\bar \phi }\dot\epsilon _m
	\big).}
\end{align}
\vspace{-0.25em}

\begin{multicols}{2}
The terms accompanying $ \mathcal R_c $ on the right hand side all involves $ \epsilon _m $ or time-derivative thereof! As such, following the reasoning in Ref. \cite{Motohashi:2015pra} , in the full Horndeski theory, all the terms involving $ \epsilon _m $ above vanishes in the superhorizon limit. On the other hand, for terms involving $ \dot \epsilon _m $, because in this theory, $ \epsilon _m \sim (k/aH)^2$, we find after a short calculation that $ \dot \epsilon _m \sim (k/aH)^2 $ as well. Physically, if $ \epsilon _m $ is to vanish in the superhorizon limit (during which primordial cosmological perturbations are ``outside'' the horizon) and if it is to hold for sufficiently long amount of time, its time derivative should be nearly zero if not vanishing as well. It is worth noting that the relation above between $ \widehat {\mathcal R}_c $ and $ \mathcal R_c $ holds for any gauge. We conclude that under the general extended disformal transformation given by (\ref{genextdt}) yielding the transformed equation (\ref{Rcsimp}) for the gauge-invariant curvature perturbation to linear order, within the framework of the full Horndeski theory, $ \widehat {\mathcal R}_c = \mathcal R_c$ in the superhorizon limit.

\bigskip
\bigskip
\section{Limitation: Invertibility of the General Extended Disformal Transformation}
\label{limigextdt}

The usefulness of the formula (\ref{Rcsimp}) {involving the curvature perturbation $ \widehat {\mathcal R}_c $ in relation to $ \mathcal R $},  hinges upon whether the transformation (\ref{genextdt}) {} is invertible or not. { The idea is that the transformation properties of the perturbations may be captured within the framework of the original theory if the transformed action is effectively ``equivalent'' (in terms of the number of physical degrees of freedom) under some conditions, to the former; for this, the disformal transformation has to be invertible. In this section,} we would like to investigate   the condition for invertibility of the transformation (\ref{genextdt}).

{We recall that} the Horndeski action is form-invariant under the special disformal transformation wherein both the conformal and disformal factors  $ (A,B) $ are functions of $ \phi  $ alone. When the dependence on $ X $ is included in these factors, the resulting action leads to equations of motion that go beyond second order in the derivatives. Nevertheless, the number of the true propagating degrees of freedom remains the same as in the original theory \cite{Gleyzes:2014qga,Gleyzes:2014dya}. We now know that this holds when the disformal transformation is invertible \cite{BenAchour:2016fzp}.

Let us revisit the disformal transformation involving $ A(\phi ,X) $ and $ B(\phi ,X) $, to see how introducing additional dependency can affect its invertibility. We have
\begin{align}
	\widehat g_{\mu \nu } 
	= 
	A(\phi ,X)g_{\mu \nu } 
	+ 
	B(\phi ,X)\phi _{;\mu }\phi _{;\nu }.
\end{align}
The equation relating $ \widehat X $ to $ X $ given by 
\begin{align}
	\widehat X = \frac{X}{A - 2BX},
\end{align}
tells us that $ \widehat X = \widehat X(\phi ,X) $ and that we can express $ X $ in terms of $ (\phi ,\widehat X) $ if 
\begin{align}
	\frac{\partial \widehat X}{\partial X} 
	=
	\frac{A - A_X X + 2B_XX^2}{(A - 2BX)^2}
	\ne 0. 
\end{align}
If this condition holds, we can express $ g_{\mu \nu } $ in terms of the hatted quantities namely, $ \widehat g_{\mu \nu }, \widehat X $, and $ \widehat \phi  = \phi  $; that is, we can invert the disformal transformation above as
\begin{align}
	g_{\mu \nu }
	&=
	\widehat A(\phi ,\widehat X)\widehat g_{\mu \nu }
	+
	\widehat B(\phi ,\widehat X)\phi _{;\mu }\phi _{;\nu },
\end{align}
where 
\begin{align}
	\widehat A(\phi ,\widehat X) 
	&= 
	\frac{1}{A(\phi ,X)},
	\nonumber
	\\[0.5em]
	\widehat B(\phi ,\widehat X)
	&=
	-\frac{B(\phi ,X)}{A(\phi ,X)}.
\end{align}

Now, suppose we add an additional dependency in the form $ Y = \phi ^{;\alpha }X_{;\alpha } $ in both $ (A,B) $. The form $ \widehat X $ is exactly of the same form as in the previous one above. But because of the added dependency, $ \widehat X = \widehat X (\phi ,X,Y) $. For the hatted $ Y $ we have
\begin{align}
	\widehat Y
	&=
	\widehat g^{\mu \nu }\phi _{;\mu} \widehat X_{;\nu},
	\nonumber
	\\[0.5em]
	\widehat Y
	&=
	\frac{1}{A}\bigg(
		g^{\mu \nu }
		-
		\frac{B\phi ^{;\mu} \phi ^{;\nu} }{A - 2BX} 
	\bigg)\phi _\mu \bigg(
		\frac{X}{A - 2BX}
	\bigg)_{;\nu}.
\end{align}
When the covariant derivative with respect to $ x^\nu $ hits $ A $ or $ B $ on the right hand most factor, in combination with other quantities in the equation above, the following dependencies are generated namely,
\begin{align}
	Z = X^{;\mu }X_{;\mu }
	\quad \text{and} \quad 
	Z_2 = \phi ^{;\alpha} \phi ^{;\mu} \phi ^{;\nu} \phi _{;\alpha \mu \nu };
\end{align}
these are in addition to $ (\phi ,X,Y) $. 

Because of these extra dependencies, the disformal transformation involving $ A(\phi ,X,Y)$ and  $ B(\phi ,X,Y) $ is invertible only when these extra dependencies are negligibly small\footnote{Even in the limit when $ B = 0 $,  the ``conformal'' transformation, $ g_{\mu \nu } \rightarrow A(\phi ,X,Y)g_{\mu \nu }, $ is no longer invertible unless the terms involving $ Z $ and $ Z_2 $ in the expression for $ {\widehat Y} $ are negligible.}. In fact, if for some energy scale, we may neglect $ Z $ and $ Z_2 $, then we may express $ \widehat Y = \widehat Y(\phi ,X,Y) $ as
\begin{align}
	\widehat Y
	&=
	\frac{2X^2(A_\phi - 2B_\phi X) + Y(A - A_XX + 2B_XX^2)}{(A - 2BX)^3}.
\end{align}
In combination with the equation for $ \widehat X = \widehat X(\phi ,X,Y) $, we may solve for
\begin{align}
	X = X(\phi ,\widehat X, \widehat Y)
	\quad \text{and} \quad 
	Y = Y(\phi ,\widehat X, \widehat Y),
\end{align}
provided the determinant of the Jacobian does not vanish:
\begin{align}
	\left|
	\begin{array}{cc}
		\partial _X\widehat X & \partial _Y\widehat X
		\\[0.35em]
		\partial _X\widehat Y & \partial _Y\widehat Y
	\end{array}
	\right|
	\ne 
	0.
\end{align}
In this case, the disformal transformation given by 
\begin{align}
	\widehat g_{\mu \nu } 
	= 
	A(\phi ,X, Y)g_{\mu \nu } 
	+ 
	B(\phi ,X, Y)\phi _{;\mu }\phi _{;\nu }
\end{align}
may be inverted as before, namely, 
\begin{align}
	g_{\mu \nu }
	&=
	\widehat A(\phi ,\widehat X, \widehat Y)\widehat g_{\mu \nu }
	+
	\widehat B(\phi ,\widehat X, \widehat Y)\phi _{;\mu }\phi _{;\nu },
\end{align}
where 
\begin{align}
	\widehat A(\phi ,\widehat X, \widehat Y) 
	&= 
	\frac{1}{A(\phi ,X, Y)},
	\nonumber
	\\[0.5em]
	\widehat B(\phi ,\widehat X, \widehat Y)
	&=
	-\frac{B(\phi ,X, Y)}{A(\phi ,X, Y)}.
\end{align}

For the general extended disformal transformation given by (\ref{genextdt}) we have a dependency on $ Z $, in addition to $ (\phi ,X,Y) $;
\begin{align}
	\widehat g_{\mu \nu }
	&=
	A(\phi ,X,Y,Z)g_{\mu \nu }
	+
	C^2(\phi ,X,Y,Z)\phi _{;\mu }\phi _{;\nu }
	\nonumber
	\\[0.5em]
	&\quad
	+\,
	CD(\phi _{;\mu }X_{;\nu } + \phi _{;\nu }X_{;\mu })
	\nonumber
	\\[0.35em]
	&\quad
	+\,
	D^2(\phi ,X,Y,Z)X_{;\mu }X_{;\nu }.
\end{align}
Needless to say, owing to its more general yet more complex nature, the transformation is, in general, not invertible. We find from their respective definitions given by (\ref{defXYZ}) and the metric inverse given by 
\begin{align}
	\widehat g^{\mu \nu }
	&=
	\frac{g^{\mu \nu }}{A}
	-
	\frac{\Phi ^{\mu} \Phi ^{\nu} }{A(A - 2\chi)} ;
	\quad
	(\chi \equiv -\tfrac{1}{2} g^{\mu \nu }\Phi _{\mu} \Phi _{\nu} ),
\end{align} 
that
\begin{align}
	\label{xtratrms}
	\widehat X
	&=
	\widehat X(\phi ,X,Y,Z),
	\\[0.35em]
	\widehat Y
	&=
	\widehat Y(\phi ,X,Y,Z,
		\underline{Z_2, \phi ^\mu Z_{;\mu},
			X^{;\mu} Y_{;\mu}, X^{;\mu} Z_{;\mu}
		}
	),
	\nonumber
	\\[0.35em]
	\widehat Z
	&=
	\widehat Z(\phi ,X,Y,Z,
		\nonumber
		\\[0.5em]
		&\qquad\quad
		\underline{Z_2, \phi ^{;\mu} Z_{;\mu}, X^{;\mu} Y_{;\mu}, 
			X^{;\mu} Z_{;\mu}, Y^{;\mu} Y_{;\mu}, Y^{;\mu} Z_{;\mu}, 
			Z^{;\mu} Z_{;\mu}
		}
	).
	\nonumber
\end{align}
The extra dependencies above (underlined) are generated due to the dependence of $ Y $ and $ Z $ on the covariant derivative of $ X $. Because of this, their hatted versions involve the derivative of $ \widehat X $, which as we see above, involves $ (\phi ,X,Y,Z) $. The derivatives of the last two functions in this dependency list, namely, $ (Y,Z) $, results in the extra dependencies. See \ref{appendB} for the explicit forms of $ \widehat X, \widehat Y, $ and $ \widehat Z $.

Admittedly, part of the limitation of this work is the non-invertibility of the general extended disformal transformation under the most general framework wherein the factors $ (A,C,D) $ depend on $ (\phi ,X,Y,Z) $. {However, we may notice that the extra terms in (\ref{xtratrms}) all involve third order derivatives of $ \phi $. This is in addition to combinations involving first and second order derivatives. For instance, 
\begin{align}
	\label{phimuZmu}
	X ^{;\mu} Z_{;\mu} 
	&= 
	-2\phi _{;\alpha }\phi _{;\beta }
	\phi ^{;\alpha \mu }\phi ^{;\beta \nu }\big(
		{\phi ^{;\rho }}_\mu \phi _{;\rho \nu }
		+
		\phi ^{;\rho }\phi _{;\rho \nu \mu }
	\big).
\end{align} 
If we \textit{assume} that there exists an energy scale wherein the third order derivatives (\ref{xtratrms}) are effectively negligible, then we will be left with the extra terms containing only up to second order derivatives; for instance, $X ^{;\mu} Z_{;\mu} \approx -2\phi _{;\alpha }\phi _{;\beta }\phi ^{;\alpha \mu }\phi ^{;\beta \nu }{\phi ^{;\rho }}_\mu \phi _{;\rho \nu }$. If these remaining terms can be neglected, then the transformation may be inverted.}

{In practice, the order-of-magnitude estimate of the size of the extra terms in (\ref{xtratrms}), sans the third order derivative terms, are better implemented within a specific inflation model that may accommodate or contain them, under the general Horndeski framework or extensions thereof. However, it may be good to have some insight about their size, be it a rough one, within the general context of slow-roll inflation in which this paper belongs. Consider for instance, $ X ^{\mu }Z_{\mu } $ given by (\ref{phimuZmu}). Given the slow-roll parameter\footnote{In canonical single-field inflation, $ \epsilon = \dot \phi ^2/2H^2 $. For general inflation within the framework of Horndeski theory, $ \epsilon  $ takes a more complicated form but $ \epsilon $ remains proportional to $\dot \phi ^2/H^2$ \cite{Kamada:2012se,Mishima:2019vlh}.} $ \epsilon \equiv -\dot H/H^2 \propto \dot \phi ^2/H^2 $ and assuming that its time derivative is progressively small\cite{Martin:2013uma} as $ \dd^n\epsilon /\dt^n \sim \mathcal O(\epsilon ^{n+1}) $, then in the unitary gauge, $ X \sim \dot \phi ^2 $ and $ Z \sim \dot X^2 $, giving us $ X^{;\mu }Z_{;\mu } \sim \mathcal O(\epsilon^6)$. Following the same logic for the other extra terms we find\footnote{$ Z_2 $ is excluded in the list because it is third order in the derivative.}
\begin{align}
	\phi ^{;\mu}Z_{;\mu }
	&\sim
	\mathcal O(\epsilon ^{5 + 1/2}),
	\,
	X^{;\mu }Z_{;\mu }
	\sim
	\mathcal O(\epsilon ^7),
	\,
	Z^{;\mu }Z_{;\mu }
	\sim 
	\mathcal O(\epsilon ^{10})
	\nonumber
	\\[0.5em]
	X^{;\mu}Y_{;\mu }
	&\sim
	\mathcal O(\epsilon ^{5 + 1/2}),	
	\,
	Y^{;\mu }Y_{;\mu }
	\sim \mathcal O(\epsilon ^7),
	\nonumber
	\\[0.5em]
	Y^{;\mu }Z_{;\mu }
	&\sim \mathcal O(\epsilon ^{8 + 1/2}).	
\end{align}
We find that in this approximation, the extra terms sans the third order derivatives, are all higher order in the slow-roll parameter than $ X\sim \mathcal O(\epsilon ),\,Y\sim \mathcal O(\epsilon ^{2 + 1/2}) $ and $ Z\sim \mathcal O(\epsilon ^4) $. For purposes of this study, then, we hold onto the \textit{assumption} that the extra terms in (\ref{xtratrms}) can be neglected. We leave it to our future studies the more in-depth investigation of these terms. 
}

Under {the mentioned assumption}, we may write the approximate system of equations for $ (\widehat X, \widehat Y, \widehat Z) $ in relation to $ (\phi ,X,Y,Z) $.
\begin{align}
	\widehat X
	&=
	\widehat X(\phi ,X,Y,Z),
	\nonumber
	\\[0.35em]
	\widehat Y
	&=
	\widehat Y(\phi ,X,Y,Z),
	\nonumber
	\\[0.35em]
	\widehat Z
	&=
	\widehat Z(\phi ,X,Y,Z).
\end{align}
The inverse relation may be found if the Jacobian is not singular; that is,
\begin{align}
	\left|
	\begin{array}{ccc}
		\partial _X\widehat X
		&
		\partial _Y \widehat X
		&
		\partial _Z \widehat X
		\nonumber
		\\[0.35em]
		\partial _X\widehat Y
		&
		\partial _Y \widehat Y
		&
		\partial _Z \widehat Y
		\nonumber
		\\[0.35em]
		\partial _X\widehat Z
		&
		\partial _Y \widehat Z
		&
		\partial _Z \widehat Z		
	\end{array}
	\right|
	\ne 
	0
\end{align}
Consequently, the inverse general extended disformal transformation may be written as
\begin{align}
     \label{eq:inversegeneralextendeddisformaltransformation}
	g_{\mu \nu }
	&=
	\widehat A g_{\mu \nu }
	+
	\widehat C^2\phi _\mu \phi _\nu 
	+
	\widehat C\widehat D(\phi _\mu X_\nu + \phi _\nu X_\mu )
	\nonumber
	\\[0.5em]
	&\qquad	
	+\,
	\widehat D^2 X_\mu X_\nu,
\end{align}
where
\begin{align}
	\widehat A(\phi ,\widehat X,\widehat Y,\widehat Z)
	&=
	\frac{1}{A(\phi ,X,Y,Z)},
	\nonumber
	\\[0.5em]
	\widehat C^2(\phi ,\widehat X,\widehat Y,\widehat Z)
	&=
	-\frac{C^2(\phi ,X,Y,Z)}{A(\phi ,X,Y,Z)},
	\nonumber
	\\[0.5em]
	\widehat D^2(\phi ,\widehat X,\widehat Y,\widehat Z)
	&=
	-\frac{D^2(\phi ,X,Y,Z)}{A(\phi ,X,Y,Z)}.
\end{align}

To sum up,  although our formula (\ref{Rcsimp}) looks concise and is suggestive enough, its  application is restricted to physical  circumstances  in which third order derivatives of $\phi$ { and other combinations of the derivatives of $ \phi $ outside the set $ \{X,Y,Z\} $}, are negligibly small. {Furthermore,} the inverse transformation is given by (\ref{eq:inversegeneralextendeddisformaltransformation}). { These conditions justify the reasoning and calculation in Sec. \ref{sec2} and \ref{sec3} involving the primordial cosmological perturbations within the context of the original Horndeski action. Here, $ A, C, $ and $ D $ are in general, dependent on $ (X,Y,Z) $ which contain at most  second-order derivatives of $\phi$. Needless to say, the extra derivative terms identified in (\ref{xtratrms}) are absent in the mentioned sections. The contents of Sec. 2 and 3 have definite meaning as a generalization of previous authors' work by including  $X$, $Y$, and $Z$ into $A$, $C,$ and $D$.}

\bigskip
\bigskip
\section{Summary}
\label{conclude}
In the present paper we have introduced the general extended version of the disformal transformation, (\ref{genextdt}), of the spacetime metric and have investigated its implications in the primordial cosmological 
perturbations. Our new transformation is more general and looks more involved than the special one (\ref{DTbasic})  that has often been  discussed in literature. Nevertheless, we have pointed out a very simple fact: the cosmological tensor perturbation is invariant under (\ref{genextdt}) and the scalar (curvature) perturbation receives modification that depends only on the conformal part $A$ to linear order, as shown in (\ref{hattedR}). Since $A$ is a general functional of $\phi$, $X$, $Y$ and $Z$,  the modification of the scalar perturbation under  (\ref{genextdt}) looks still awkward to deal with in terms of $\varphi $ and $\delta \phi$.   We have, however, noted that something peculiar  happens if we rewrite the modification in terms of the gauge-invariant comoving density perturbation $\epsilon_{m}$  in  (\ref{eminem}). Equation (\ref{Rcsimp}) is the resulting transformation formula of the curvature perturbation which is a combination of $\epsilon_{m}$ and ${\dot \epsilon}_{m}$. Since $\epsilon_{m}$ and ${\dot \epsilon}_{m}$ both vanish in the superhorizon limit we can conclude that the curvature perturbation is invariant in the same limit under  (\ref{genextdt}), within the framework of the full Horndeski theory.

The usefulness of the formula (\ref{Rcsimp}) is ensured when one can go freely back and forth between two frames before and after the transformation  (\ref{genextdt}). {In line with this,} we have examined conditions for the existence of the inverse of the  general  extended disformal transformation (\ref{genextdt}).  It has been made clear  that the transformation (\ref{genextdt}) is invertible  when certain third order derivative terms third order derivatives of $\phi$ { and other combinations of the derivatives of $ \phi $ outside the set $ \{X,Y,Z\} $}, are negligibly small. Only in such cases our claims (a) and (b) mentioned in  Section \ref{sec1} are meaningful.

Under the physical situations  in which  the inverse transformation (\ref{eq:inversegeneralextendeddisformaltransformation}) exists, the invariance of the curvature perturbation on the superhorizon scale is  so nice a property that we cannot resist contemplating and searching for a broader class of scalar-tensor theory with the help of  (\ref{genextdt}) beyond what has been considered in literature. On applying the transformation (\ref{genextdt}), however, we immediately encounter higher derivative terms of the scalar field $\phi$, and the central concern here is how to avoid the Ostrogradsky instability. Although solving the stability problem is beyond the scope of the present paper, we have collected  the transformation formula of the Ricci tensor, scalar curvature, and the derivatives of $\phi$ in \ref{appendA} in order to get a glimpse of generalized scalar-tensor theories. As discussed there, it is extremely curious that some of the third-order derivative terms involving $\phi$, $X$ and $\chi \equiv -\frac{1}{2} g^{\mu \nu}\Phi_{\mu}\Phi_{\nu}$,  when  taken as independent fields, are put into the form of second-order derivatives. Although we cannot say anything definite any further at present, this gives us a strong impetus to search for more general scalar-tensor theories.   

{For our future studies, we would like to dig deeper into the possibility of including higher order derivative terms in the disformal transformation while preserving invertibility. A possible area to pursue may include a multi-field scenario, hints of which we can find in the appendix. Furthermore, we may also consider a disformal transformation wherein there is a separate transformation rule for the ``scalar'' field to possibly offset the extra higher order derivative terms.}

\section*{Acknowledgments}
This work is funded by the UP System Enhanced Creative Work and Research Grant (ECWRG 2018-02-11). AL Alinea would like to also acknowledge Osaka University (Japan) where some parts of this work were initiated, and the Ministry of Education (Japan) for the scholarship grant for his stay in the said university. Last but not least, the authors would like to thank
the anonymous referee, who kindly provided them with many important suggestions to improve the manuscript.

\bigskip 
\bigskip
\appendix
\section{The Transformation of the Horndeski Action}
\label{appendA}

The concise formula (\ref{Rcsimp}) together with the vanishing of 
$ \widehat {\mathcal R}_c - \mathcal R_c$  in the superhorizon limit motivates us to go further into a search for a general class of scalar-tensor theories by using the transformation (\ref{genextdt}).  Here we examine transformation properties of the Ricci tensor, scalar curvature and the derivatives of the scalar field in order to see what the general class of theories would look like,  although the stability 
problem has to be  left open.

\bigskip

\subsection{Toward a more general class of scalar-tensor theories}

The Horndeski action consists of four sub-Lagrangians concomitant of the combinations of metric, a scalar field, and derivatives thereof, yielding second order equation of motion \cite{Horndeski:1974wa}. Following current literature's notation we may write the action as
\begin{align}
	S
	&=
	\int \dd^4x\sqrt{-g} \left(
		\mathcal L_2 + \mathcal L_3 + \mathcal L_4 + \mathcal L_5
	\right),
\end{align}
where
\begin{align}
	\mathcal L_2
	&=
	P,
	\qquad
	\mathcal L_3
	=
	-G_3\,\square\phi,
	\\[0.5em]
	\mathcal L_4
	&=
	G_4R
	+ 
	G_{4,X}[(\square \phi )^2 - \phi _{;\mu \nu }\phi ^{;\mu \nu }],
	\nonumber
	\\[0.5em]
	\mathcal L_5
	&=
	G_5G_{\mu \nu }\phi ^{;\mu \nu }
	\nonumber
	\\[0.5em]
	&
	-\,
	\frac{1}{3!}G_{5,X}[
	(\square\phi )^3
	-
	3(\square\phi )\phi _{;\mu \nu }\phi ^{;\mu \nu }
	+
	2\phi _{;\mu \alpha }\phi ^{;\alpha \nu }\phi ^{;\mu }{}_{\nu }
	],
	\nonumber
\end{align}
with $ \square \phi = g^{\mu \nu }\phi _{;\mu \nu } $, $ R_{\mu \nu }$ is the Ricci tensor, $ R $ is the scalar curvature$, G_{\mu \nu } $ is the Einstein tensor, and  the functionals $ P,\, G_3,\, G_4, $ and $ G_5 $, generally depend on $ (X, \phi ) $. In the pursuit of formulating viable theories of dark energy and addressing other unexplained puzzles beyond what could be satisfactorily accounted for by general relativity in cosmology, the Horndeski action, many of its extensions, and hybrid formulations, have been actively studied \cite{Kobayashi:2019hrl} and are still being pursued by serious researchers in the field. Nowadays, we know that $ \mathcal L_4 $ and $ \mathcal L_5 $ are strongly constrained \cite	{Lombriser:2015sxa,Bettoni:2016mij,Creminelli:2017sry,Sakstein:2017xjx,Ezquiaga:2017ekz} based on the measurement of the gravitational wave signal GW170817 \cite{GBM:2017lvd,TheLIGOScientific:2017qsa,Kase:2018aps}.

Having said this, we take our freedom to present some insights into the transformation of the Horndeski action under our version of extended disformal transformation. In so doing, the intent is not necessarily to provide light into a new and more satisfactory theory overshadowing the present content. In fact as we shall see, the transformed Horndeski action has instability issues; in the absence of suitable constraint(s) or conditions, the undesirable mode(s) corresponding to the presence of ghost(s) may be inevitable. Having established however, that the primordial cosmological perturbations are invariant in the superhorizon limit at least within the framework of the full Horndeski theory (the main object of this work), we are driven by curiosity to at least have a peek on what this transformation could do on this theory. We leave investigations to take a deeper look into this matter for future studies. 

\bigskip
\subsection{The Ricci tensor under the transformation}

The Horndeski action with its generality far overshadows the simplicity of a single-term Lagrangian in the Einstein-Hilbert action (excluding the matter term). Even under the special disformal transformation, $ g_{\mu \nu } \rightarrow \widehat g_{\mu \nu } = A(\phi )g_{\mu \nu } + B(\phi )\phi _{;\mu }\phi _{;\nu }$, under which the Horndeski structure is preserved, the resulting expressions for the terms in this action such as the Ricci tensor and scalar curvature can be quite messy. With our modest aim in this section, we focus our attention to the extended disformal transformation of the form given by
\begin{align}
	\label{limgenextDT}
	&g_{\mu \nu } 
	\rightarrow 
	\widehat g_{\mu \nu } 
	= 
	A(\phi )g_{\mu \nu } + B(\phi )\Phi _{;\mu }\Phi_{;\nu },
	\\[0.5em]
	&
	\text{where} 
	\quad
	\Phi _\mu = C(\phi )\phi _{\mu } + D(\phi )X_{\mu }.
	\nonumber
\end{align}
This is a special case of (\ref{genextdt}) where $ A,B,C, $ and $ D $ all depends on $ (\phi ,X,Y,Z) $. The inverse of the disformal metric follows from the Sherman-Morrison formula.
\begin{align}
	\widehat g^{\mu \nu }
	&=
	E_2g^{\mu \nu }
	-
	E_1E_2\Phi ^\mu \Phi ^\nu,
\end{align}
Here, $\chi \equiv -\tfrac{1}{2}\,g^{\mu \nu }\Phi _\mu \Phi _\nu, \, E_1 \equiv 1/(A - 2\chi ),$ and $ E_2 \equiv 1/A $. The quantity $ \chi  $ is a generalized expression for $ X $ in the Horndeski action. 

With the disformal metric and its inverse at hand, we now proceed to derive the disformal Ricci tensor $ \widehat R_{\mu \nu } $ from the expression for the disformal Riemann tensor $ \widehat R^\alpha {}_{\mu \beta \nu } $. The latter is given by a familiar equation involving Christoffel symbols, $ \widehat \Gamma ^\alpha _{\mu \nu } $, now wearing a hat. We have
\begin{align}
	\widehat R^\alpha {}_{\mu \beta \nu }
	&=
	-\widehat \Gamma ^{\alpha }_{\mu \beta ,\nu }
	+
	\widehat \Gamma ^\alpha _{\mu \nu ,\beta }
	-
	\widehat \Gamma ^\rho _{\mu \beta }\widehat \Gamma ^\alpha _{\rho \nu }
	+
	\widehat \Gamma ^\rho _{\mu \nu }\widehat \Gamma ^\alpha _{\rho \beta },
\end{align}
where
\begin{align}
	\widehat \Gamma ^\alpha _{\mu \nu }
	&=
	\frac{1}{2} \widehat g^{\alpha \beta }\left(
		\widehat g_{\beta \mu ,\nu }
		+
		\widehat g_{\nu \beta ,\mu }
		-
		\widehat g_{\mu \nu ,\beta }
	\right),
	\nonumber
	\\[0.5em]
	\widehat \Gamma ^\alpha _{\mu \nu }
	&=
	\Gamma ^\alpha _{\mu \nu } + C^\alpha _{\mu \nu },
\end{align}
with $C^\alpha _{\mu \nu }$ being a tensor symmetric in the two covariant indices $ (\mu, \nu ) $, defined by 
\begin{align}
	C^\alpha _{\mu \nu }
	&\equiv
	\tfrac{1}{2}E_2\big(
		A_{;\nu }\delta ^\alpha _\mu 
		+
		A_{;\mu }\delta ^\alpha _\nu 
		-
		A^{;\alpha }g_{\mu \nu }	
	\big)
	\nonumber
	\\[0.5em]
	&\quad 
	-\,
	\tfrac{1}{2}E_1E_2\Phi ^\alpha\big(
			A _{;\nu }\Phi _\mu 
			+
			A _{;\mu }\Phi _\nu 
			-
			A _{;\beta }\Phi ^\beta g_{\mu \nu }
	\big)
	\nonumber
	\\[0.5em]
	&\quad 
	+\,
	\big(
		E_2g^{\alpha \beta }
		-
		E_1E_2\Phi ^\alpha \Phi ^\beta
	\big)
	(
		\Phi _\mu\Phi _{[\beta;\nu ]} 
		+
		\Phi _{\nu }\Phi _{[\beta;\mu]} 	
	)
	\nonumber
	\\[0.5em]
	&\quad 	
	+\,
	E_1\Phi^\alpha\Phi _{(\mu;\nu )}.		
\end{align}
Pairs of indices inside a pair of square brackets and a pair of parentheses indicate antisymmetric and symmetric combinations respectively, namely, $ \Phi _{[\mu ;\nu ]} = \tfrac{1}{2}\,(\Phi _{\mu ;\nu } - \Phi _{\nu ;\mu }) $ and $ \Phi _{(\mu ;\nu )} = \tfrac{1}{2}\, (\Phi _{\mu ;\nu } + \Phi _{;\nu ;\mu })$. Substitution from the equation above for $ C^\alpha _{\mu \nu } $ in the equation for $ \widehat \Gamma ^\alpha _{\mu \nu } $ and using the result in the relation for the hatted Riemann tensor, we find upon contracting the indices $ \alpha  $ and $ \beta  $ a rather lengthy expression for the Ricci tensor.
\end{multicols}
\begin{align}
	\widehat R_{\mu \nu }
	&=
	R_{\mu \nu } 
	+
	E_1 \Phi _{(\mu ;\nu) \alpha }\Phi ^{\alpha }
	+
	E_1\Phi _{(\mu ;\nu )}
	\big[
		{\Phi _\alpha }^{;\alpha }
		+ 
		E_1(\Phi ^{\alpha }\chi_{;\alpha } - A'E_2\chi\phi ^{;\alpha }\Phi _\alpha )
	\big]
	\nonumber
	\\[0.5em]
	&\quad
	-\,
	\tfrac{1}{2}E_1E_2^2\,\Phi _\mu \Phi _\nu \big[
		\Phi _{\alpha; \beta }\Phi ^\beta (
			\Phi ^{\alpha; \rho }\Phi _\rho 
			+ 
			2\chi^{;\alpha} 
			- 
			2A'\phi ^{;\alpha} 
		)
		+
		\chi^{;\alpha }\chi_{;\alpha }
		-
		2A'( 
			A'X
			+
			\phi ^{;\alpha }\chi_{;\alpha }
		)
	\big]
	\nonumber
	\\[0.5em]
	&\quad
	+\,
	\tfrac{1}{2} E_1E_2
	\Phi _{\alpha ;(\mu }\Phi _{\nu) ;\beta }
	\Phi ^\alpha \Phi ^\beta 
	+
	\tfrac{1}{2} E_1E_2\Phi _{\mu ;\alpha }\Phi _{\nu ;\beta }
	\Phi ^\alpha \Phi ^\beta 
	+
	\tfrac{1}{2} E_2\Phi _\mu \big\{
		E_1\Phi _{\nu ;\alpha \beta }\Phi ^\alpha \Phi ^\beta 
		\nonumber
		\\[0.5em]
		&\quad\quad
		+\,
		E_1\chi_{;\nu \alpha }\Phi ^\alpha 
		+
		E_1\Phi _{\nu ;\alpha }\Phi ^\alpha [
			{\Phi _\alpha }^{;\alpha }
			-
			A'E_1E_2\phi ^{;\beta }\Phi _\beta (A - \chi)
			+
			E_1\Phi ^\alpha \chi_{;\alpha }
		]
		\nonumber
		\\[0.5em]
		&\quad\quad
		+\,
		A'E_1E_2(A - \chi)\Phi _{\nu ;\alpha }\phi ^{;\alpha }
		+
		E_1\Phi _{\alpha ;\nu }(A'E_2 \chi\phi ^{;\alpha } - \chi^{;\alpha })
		-
		A'E_1\phi _{;\alpha \nu }\Phi ^\alpha 
		\nonumber
		\\[0.5em]
		&\quad\quad
		+\,
		(
			{\Phi ^\alpha }_{;\nu \alpha }
			-
			{{\Phi _\nu }^{;\alpha }}_{\alpha }
		)
	\big\}
	+
	\big\{(\mu \leftrightarrow \nu )\big\}
	\nonumber
	\\[0.5em]
	&\quad
	+\,
	\tfrac{1}{4}E_1E_2^2 \Phi ^\alpha \big[
		\chi_{;\mu }(3A\Phi _{\nu ;\alpha } + 2\chi\Phi _{\alpha ;\nu })
		+
		\chi_{;\nu }(3A\Phi _{\mu ;\alpha } + 2\chi\Phi _{\alpha ;\mu })
	\big]
	\nonumber
	\\[0.5em]
	&\quad
	-\,
	\tfrac{1}{4} A'E_1E_2\big[
		\phi _{;\mu }(2\Phi _{\nu ;\alpha } + \Phi _{\alpha ;\nu })
		+
		\phi _{;\nu }(2\Phi _{\mu ;\alpha } + \Phi _{\alpha ;\mu })
	\big]\Phi ^\alpha 
	+
	E_2(\Phi _{\alpha ;(\mu }{\Phi _{\nu) }}^{;\alpha })
	\nonumber
	\\[0.5em]
	&\quad
	-\,
	E_1E_2(A - \chi)\Phi _{\mu ;\alpha }{\Phi _\nu }^{;\alpha }
	+
	E_1E_2\Phi _{(\mu }\chi_{;\nu )}\big[
		{\Phi _\alpha }^{;\alpha }
		-
		A'E_1E_2\phi ^{;\alpha }\Phi _\alpha (A - \chi)
		+
		E_1\Phi ^\alpha \chi_{;\alpha }
	\big]
	\nonumber
	\\[0.5em]
	&\quad
	-\,
	\tfrac{1}{2} A'E_1^2E_2(5A - 6\chi)
	\phi _{;(\mu }\chi_{;\nu )}
	+
	\tfrac{1}{4} E_1^2E_2^2\big[7A^2 - 4\chi(A + \chi)\big]
	\chi_{;\mu }\chi_{;\nu }
	\nonumber
	\\[0.5em]
	&\quad
	-\,
	E_1E_2\phi _{;(\mu }\Phi _{\nu)} \big\{
		E_1E_2\phi ^{;\alpha }\Phi _\alpha [
			AA''(A - 2\chi)
			-
			A'^2(A - \chi)
		]
		+
		A'{\Phi _\alpha }^{;\alpha }
		+
		A'E_1\Phi _\alpha \chi^{;\alpha }
	\big\}
	\nonumber
	\\[0.5em]
	&\quad
	+\,
	\tfrac{1}{2} E_2 g_{\mu \nu }\big\{
		A'E_1\Phi ^\alpha (\phi _{;\beta }\Phi ^\beta )_{;\alpha }
		+
		E_1^2E_2(\phi _{;\alpha }\Phi ^\alpha )^2[
			AA''(A - 2\chi)
			-
			A'^2(A - \chi)
		]
		\nonumber
		\\[0.5em]
		&\quad\quad
		+\,
		2E_1E_2X[
			AA''(A - 2\chi) + A'^2\chi
		]
		+
		A'E_1\,\phi _{;\alpha }\Phi ^\alpha \,
		{\Phi _\beta }^{;\beta }
		-
		A'\square \phi 
		\nonumber
		\\[0.5em]
		&\quad\quad		
		+\,
		A'E_1^2\,\phi _{;\alpha }\Phi ^\alpha \,\Phi _\beta \chi^{;\beta }
		+
		A'E_1\phi _{;\alpha }\chi^{;\alpha }
	\big\}
	\nonumber
	\\[0.5em]
	&\quad 
	+\,
	\tfrac{1}{2} E_1^2E_2^2\phi _{;\mu }\phi _{;\nu }\big[
		2A\chi(3AA'' - 4A'^2)
		-
		(A^2 + 2\chi^2)(2AA'' - 3A'^2)
	\big]	
	\nonumber
	\\[0.5em]
	&\quad
	+\,
	E_1\chi_{;\mu \nu }
	-
	E_1E_2A'(A - \chi)\phi _{;\mu \nu }
	+
	E_1E_2\chi{\Phi ^\alpha }_{;\nu }\Phi _{\alpha ;\mu }
	-
	\tfrac{1}{4}E_2^2\Phi _{\alpha ;\mu }
	\Phi _{\beta ;\nu }\Phi ^\alpha \Phi ^\beta.
\end{align}
\begin{multicols}{2}
Here, the prime indicates differentiation with respect to the field $ \phi  $. The expression on the right hand side is symmetric with respect to the indices $(\mu,\nu)$ confirming the symmetry of $\widehat R_{\mu\nu}$. Furthermore, we have checked\footnote{The derivation of this equation is done by hand and checked using a Computer Algebra System.} that the equation above for the hatted Ricci tensor correctly reduces to that in Ref. \cite{Bettoni:2013diz} in the limit where $ \widehat g_{\mu \nu } = A(\phi )g_{\mu \nu } + B(\phi )\phi _{;\mu }\phi _{;\nu } $, saved for some unintentional typographical errors in this reference and slight difference in notation.

The disformally transformed Ricci tensor is a sum of the original Ricci tensor and terms involving $ A $ and scalar combinations of $ \Phi _\mu $, its derivatives, and other derivative terms. Needless to say, when the disformal term in the transformation given by (\ref{limgenextDT}) vanishes and $ A = 1 $, all these additional terms vanish. Note that in the expression above for $ \widehat R_{\mu \nu} $, if we assume that $ \phi, X $, and $ \chi $ are all independent scalar fields, some terms on the right hand side are suggestively third order in the derivative indices. To check this, we isolate these terms in the expression below and perform some simplification.
\begin{align}
	&E_1\,(\Phi ^\alpha \Phi _{\mu ;(\nu \alpha )} + \chi _{;\mu \nu })
	+
	\tfrac{1}{2} E_1E_2\,\Phi ^\alpha\Phi _\mu \big(
		\Phi ^\beta\,
		\Phi _{\nu ;\alpha \beta }
		+
		\chi _{;\nu \alpha }
	\big)
	\nonumber
	\\[0.5em]
	&\quad 
	+\,
	\tfrac{1}{2} E_1E_2\,\Phi ^\alpha\Phi _\nu  \big(
		\Phi ^\beta\,\Phi _{\mu ;\alpha \beta }
		+
		\chi _{;\mu \alpha }
	\big)
	\nonumber
	\\[0.5em]
	&\quad 
	-\,
	\tfrac{1}{2}E_2\,\Phi _\mu \big(
		{{\Phi _{\nu}}^{;\alpha }}_\alpha 
		-
		{\Phi ^\alpha }_{;\nu \alpha }
	\big)
	\nonumber
	\\[0.5em]
	&\quad 
	-\,
	\tfrac{1}{2}E_2\,\Phi _\nu\big(
		{{\Phi _{\mu}}^{;\alpha }}_\alpha 
		-
		{\Phi ^\alpha }_{;\mu \alpha }
	\big).
\end{align}
Noting that $\Phi _\mu \equiv C(\phi )\,\phi _{;\mu } + D(\phi )\,X_{;\mu }$ we find 
\begin{align}
	\Phi _{\mu ;\nu }
	&=
	\Phi _{\nu ;\mu }
	-
	\tilde S _{\mu \nu },
	\nonumber
	\\[0.5em]
	\text{where}\quad 
	\tilde S _{\mu \nu }
	&\equiv
	D'(\phi )\,(
		\phi _{;\mu }\,X_{;\nu }
		-
		\phi _{;\nu}\,X_{;\mu }
	).
\end{align}
By further using the definition of the Riemann curvature tensor in terms of covariant derivatives, we can express all of the above ``third-order looking'' terms as
\end{multicols}
\begin{align}
	\label{3rdorderlookingterms}
	E_1\,(\Phi ^\alpha \Phi _{\mu ;(\nu \alpha )} + \chi _{;\mu \nu })
	&=
	E_1\big[
		\Phi ^\alpha \Phi ^\beta R_{\mu \beta \alpha \nu }
		-
		\Phi ^\alpha _{;\mu }\Phi _{\alpha ;\nu }
		+
		\Phi ^\alpha \big(
			\tfrac{1}{2} \tilde S_{\mu \nu ;\alpha }
			+
			 \tilde S_{\alpha \mu ;\nu }
		\big)
	\big],
	\nonumber
	\\[0.5em]
	\tfrac{1}{2} E_1E_2\,\Phi ^\alpha\Phi _\mu \big(
		\Phi ^\beta\,
		\Phi _{\nu ;\alpha \beta }
		+
		\chi _{;\nu \alpha }
	\big)	
	&=
	\tfrac{1}{2}E_1E_2\,\Phi ^\alpha \Phi _\mu \big(
		\Phi ^\beta  \tilde S_{\beta \nu; \alpha }
		-
		\Phi ^\beta _{;\nu }\Phi _{\beta ;\alpha }
	\big),
	\nonumber
	\\[0.5em]
	\tfrac{1}{2} E_1E_2\,\Phi ^\alpha\Phi _\nu  \big(
		\Phi ^\beta\,\Phi _{\mu ;\alpha \beta }
		+
		\chi _{;\mu \alpha }
	\big)	
	&=
	\tfrac{1}{2}E_1E_2\,\Phi ^\alpha \Phi _\nu \big(
		\Phi ^\beta \tilde S_{\beta \mu ;\alpha }
		-
		\Phi ^\beta _{;\mu }\Phi _{\beta ;\alpha }
	\big),
	\nonumber
	\\[0.5em]
	-
	\tfrac{1}{2}E_2\,\Phi _\mu \big(
		{{\Phi _{\nu}}^{;\alpha }}_\alpha 
		-
		{\Phi ^\alpha }_{;\nu \alpha }
	\big)	
	&=
	-\tfrac{1}{2}E_2\,\Phi _\mu { \tilde S_{\alpha \nu }}{}^{;\alpha} ,
	\nonumber
	\\[0.5em]
	-
	\tfrac{1}{2}E_2\,\Phi _\nu  \big(
		{{\Phi _{\mu}}^{;\alpha }}_\alpha 
		-
		{\Phi ^\alpha }_{;\mu \alpha }
	\big)	
	&=
	-\tfrac{1}{2}E_2\,\Phi _\nu { \tilde S_{\alpha \mu }}{}^{;\alpha}.
\end{align}
\begin{multicols}{2}
\noindent
Hence, taking $ \phi, X $, and $ \chi $ as independent scalar fields, the disformally transformed hatted Ricci tensor is only second-order in the tensor derivative indices. 

\bigskip
\subsection{The scalar curvature under the transformation}

The transformed scalar curvature can be obtained following the contraction of the hatted inverse metric and the hatted Ricci tensor given above; \textit{i.e.}, $ \widehat R = \widehat g^{\mu \nu }\widehat R_{\mu \nu}  $. We find
\end{multicols}
\vspace{-2.25em}
\begin{align}	
	\widehat R
	&=
	E_2 R 
	-
	2E_1E_2\,\Phi ^\alpha \Phi ^\beta R_{\alpha \beta }
	+ 
	\tfrac{1}{2}E_1E_2^2\,\Phi ^{\alpha ;\mu }\Phi _{\mu \,}
	\Phi _{\alpha ;\beta }\Phi ^\beta 
	-
	2E_1E_2\,\Phi ^\beta {{\tilde S}_{\alpha \beta }}^{\quad;\alpha }
	\nonumber
	\\[0.5em]
	&\quad 
	+\,
	E_1E_2^2\,\Phi _{\alpha ;\beta }\chi ^{;\alpha }\Phi ^\beta 	
	+
	E_2^2\,\Phi _{\alpha ;\beta }\Phi ^{\beta ;\alpha }	
	-
	2E_1E_2^2\,\Phi _{\alpha ;\beta }\Phi ^{\alpha ;\beta }
	(A - \chi )
	\nonumber
	\\[0.5em]
	&\quad 
	-\,
	A'(E_1E_2)^2\,\Phi _{\alpha ;\beta }\phi ^{;\alpha }\Phi ^\beta (A - \chi )	
	-
	A'(E_1E_2)^2\chi \,\phi _{;\alpha \beta }\Phi ^\alpha \Phi ^\beta 	
	\nonumber
	\\[0.5em]
	&\quad 
	+\,
	\tfrac{1}{2} E_1E_2^2\,\Phi _{\alpha ;\beta }\Phi ^\alpha \Phi ^\beta 
	\big[
		{\Phi ^\alpha }_{;\alpha } 
		- 
		A'E_1E_2\,(\phi ^{;\alpha }\Phi _\alpha )(A - \chi )
		+
		E_1\,(\Phi ^\alpha \chi _{;\alpha })
	\big]
	\nonumber
	\\[0.5em]
	&\quad 
	+\,
	\tfrac{1}{2} A'E_1E_2^3\,\Phi _{\alpha ;\beta }\Phi ^\alpha \phi ^{;\beta }
	(A - \chi )
	+
	A'(E_1E_2)^2\,\Phi ^\alpha (\phi ^{;\beta }\Phi _\beta )_{;\alpha }
	(2A - 3\chi )
	\nonumber
	\\[0.5em]
	&\quad 
	+\,
	\tfrac{1}{2}(E_1E_2)^2(\chi ^{;\alpha }\chi _{;\alpha })(5A - 2\chi )
	-
	\tfrac{1}{2} A'E_1^2E_2^3(\phi ^{;\alpha }\chi _{;\alpha })
	(A^2 + 7A\chi  - 2\chi ^2)
	\nonumber
	\\[0.5em]
	&\quad 
	-\,
	4(E_1E_2)^3\,X\chi ^2\big[
		A'^2(7A - 6\chi )
		-
		AA''(7A - 8\chi )
	\big]
	+
	E_1E_2({\Phi ^\alpha }_{;\alpha})^2	
	\nonumber
	\\[0.5em]
	&\quad 
	+\,
	4E_1^3E_2^2\,X\chi \big[
		A'^2(4A - \chi )
		-
		AA''(6A - 7\chi )
	\big]
	-
	A'E_1E_2^2(\square \phi )(3A - 4\chi )		
	\nonumber
	\\[0.5em]
	&\quad 
	-\,
	\tfrac{1}{2} E_1^2E_2^3\,(\phi ^{;\alpha }\Phi _\alpha )^2\big[
		A'^2(5A - 6\chi )
		-
		4AA''(A - 2\chi )
	\big]
	+
	\tfrac{1}{2} (E_1E_2)^2(\Phi ^\alpha \chi _{;\alpha })^2
	\nonumber
	\\[0.5em]
	&\quad 
	-\,
	E_1^3\,X\big[3A'^2  - 2A''(3A - 4\chi )\big]
	+
	A'(E_1E_2)^2(\phi ^{;\alpha }\Phi _{\alpha })
	({\Phi ^\alpha }_{;\alpha})(A - 4\chi )	
	\nonumber
	\\[0.5em]
	&\quad 
	+\,
	\tfrac{1}{2} 
	(E_1E_2)^2(\Phi ^\alpha \chi _{;\alpha })({\Phi ^\alpha }_{;\alpha})
	(5A - 2\chi )
	+
	\tfrac{1}{2} A'E_1^2E_2^3(\phi ^{;\alpha }\Phi _\alpha )
	(\Phi ^\beta \chi _{;\beta})(3A + \chi ).	
\end{align}	
\begin{multicols}{2}			
The transformed scalar curvature is a sum of the original scalar curvature multiplied by the inverse of the conformal factor, and terms involving $ A $ and scalar combinations of $ \Phi _\mu $, its derivatives, and other derivative terms. It reduces to the result in Ref. \cite{Bettoni:2013diz} in the special disformal transformation, $ \widehat g_{\mu \nu } = A(\phi )g_{\mu \nu } + B(\phi )\phi _{;\mu }\phi _{;\nu } $, saved for some unintentional typographical errors in this reference and slight difference in notation. Similar to that of the hatted Ricci tensor, the hatted scalar curvature is at most second-order in the tensor derivative indices if we take $ \phi , X, $ and $ \chi  $ as independent scalar fields. 

The form of the transformed Ricci tensor and the scalar curvature being at most second order in the tensor derivative indices when $ (\phi ,X ,\chi ) $ are considered as independent fields, may give us an idea about a multi-field approach for the transformed Horndeski action. This approach is beyond the scope of the current work. However, it is good to think of this as one possible way out to the problematic terms (leading to equations of motion involving derivatives with order higher than two) introduced by the general extended disformal transformation in the Horndeski action. The idea is somewhat analogous to $ f(R) $ theories wherein the introduction of auxiliary field(s) can simplify the action in an alternative formalism. For the case at hand, the introduction of auxiliary fields seem to hold some promise but the equations of motion remain with derivative terms of order higher than two.

\bigskip
\subsection{The derivatives of $\phi$ in the Horndeski action}

The other terms in the Horndeski action involving covariant derivatives of $ \phi  $ are straightforward to transform. For instance, we find after a short calculation that 
\begin{align}
	\phi _{;\widehat \mu \widehat \nu }
	&=
	\phi _{;\mu \nu } - C^\alpha {}_{\mu \nu }\,\phi _{;\alpha },
	\\[0.5em]
	\phi ^{;\widehat \mu }{}_{\widehat \nu }
	&=
	E_2\phi ^{;\mu }{}_{\nu }
	-
	E_1E_2\Phi ^\mu \Phi ^\alpha \phi _{;\alpha \nu }
	-
	E_2C^{\alpha  \mu }{}_\nu \phi _{;\alpha  }
	\nonumber
	\\[0.5em]
	&\quad
	+\,
	E_1E_2\Phi ^\mu \Phi ^\alpha 
	C^{\beta }{}_{\alpha \nu }\phi _{;\beta },
	\nonumber
	\\[0.5em]
	\widehat \square \phi 
	&=
	E_2\,\square \phi 
	-
	E_1E_2\Phi ^\mu \Phi ^\nu  \phi _{;\mu \nu }
	-
	E_2C^{\mu \nu }{}_\nu  \phi _{;\mu  }
	\nonumber
	\\[0.5em]
	&\quad
	+\,
	E_1E_2\Phi ^\mu \Phi ^\nu C^{\alpha }{}_{\mu \nu }\phi _{;\alpha },
	\nonumber
	\\[0.5em]
	\phi ^{;\widehat \mu \widehat \nu }
	&=
	E_2(g^{\mu \alpha } - E_1\Phi_\mu \Phi ^\alpha  )\big(
		E_2\phi ^{;\mu }{}_{\alpha  }
		-
		E_1E_2\Phi ^\mu \Phi ^\beta  \phi _{;\beta  \alpha  }
		\nonumber
		\\[0.5em]
		&\qquad
		-\,
		E_2C^{\beta   \mu }{}_\alpha  \phi _{;\beta   }
		+
		E_1E_2\Phi ^\mu \Phi ^\beta  
		C^{\rho  }{}_{\beta  \alpha }\phi _{;\rho  }
	\big),
	\nonumber
\end{align}
where the hat in `$ ;\widehat \mu $' and `$ \widehat \square $', means the covariant differentiation is taken with respect to the hatted metric $ \widehat g_{\mu \nu } $. From this set of equations, one can easily calculate terms involving $ (\widehat \square\phi )^2 - \phi ^{;\widehat \mu \widehat \nu }\phi _{;\widehat \mu \widehat \nu }$ in $ \mathcal L_4 $ and $ (\widehat \square\phi )^3	- 	3(\widehat \square\phi )\phi _{;\widehat \mu \widehat \nu }\phi ^{;\widehat \mu \widehat \nu } + 2\phi _{;\widehat \mu \widehat \alpha }\phi ^{;\widehat \alpha \widehat \nu }\phi ^{;\widehat \mu }{}_{\widehat \nu } $ in $ \mathcal L_5 $. Similar to that of the hatted Ricci tensor and hatted scalar curvature, these terms are at most second-order in the tensor derivative indices if we take $ \phi , X, $ and $ \chi  $ as independent scalar fields. 

The fundamental scalar field in the original Horndeski action is $ \phi $ accompanied by its corresponding kinetic term $ X $. Let us now hold off the assumption that $ \phi , X, $ and $ \chi $ are independent scalar fields and consider only $ \phi $ as our fundamental scalar field. Consequently, if we expand the equations above for the hatted Ricci tensor and hatted scalar curvature in terms of $ R, R_{\mu \nu }, X$, and derivatives of $ \phi  $, derivative terms involving $ \phi $ higher than two appear. In the special disformal transformation employed in Ref. \cite{Bettoni:2013diz}, there are also third-order derivative terms in the intermediate steps leading to the transformed scalar curvature and Ricci tensor, but they are assimilated by Riemann/Ricci tensor (e.g., $ X_{;\mu \nu } + \phi ^{;\alpha }\phi _{;\mu \nu \alpha } \sim -R_{\mu \alpha \nu \beta } + \cdots $). The other residual third-order derivative terms are integrated away, in combination with other transformed terms in the Horndeski action. For the case at hand, this is highly unlikely to happen as even making $ A $ or $ B $ depends on $ X $ in the transformation $ g_{\mu \nu } \rightarrow \widehat g_{\mu \nu } = A(\phi )g_{\mu \nu } + B(\phi )\phi _{;\mu }\phi _{;\nu }$, can spoil the Horndeski structure. 

For the functional derivatives in the original Horndeski action, $ G_{4,X} $ and $ G_{5,X} $, we hit a stumbling block when transforming them under (\ref{limgenextDT}). The relationship between $ \widehat X $ and $ X $ is no longer invertible as in the special disformal transformation; in particular,
\begin{align}
	\label{widehX}
	\widehat X
	&=
	\frac{
		2AX + 2D^2X\,X^{;\alpha }X_{;\alpha } + D^2(\phi ^{;\alpha }X_{;\alpha })^2
	}{
		2A(A - 2C^2X + 2CD\,\phi ^{;\alpha }X_{;\alpha } 
		+ D^2\,X^{;\alpha }X_{;\alpha } )
	}.
\end{align}
The right hand side is not a functional of $ (\phi ,X) $ alone but includes their derivatives. We cannot easily compute $ \partial X/\partial \widehat X $ given this equation only without any additional conditions or assumptions. 

In addition to this, the integral measure in the action involves derivative terms $ \phi ^{;\alpha }X_{;\alpha } $ and $ X ^{;\alpha }X_{;\alpha } $ coming from the transformation of the determinant of the metric. In particular, under (\ref{limgenextDT}) we have
\begin{align}
	\sqrt{-\widehat g}\,\dd^4x
	&=
	A^\frac{3}{2}[A
		-
		2(	C^2X - CD\,\phi ^{;\alpha }X_{;\alpha } 
		\\[0.5em]
		&\qquad\qquad\quad	
			-\,
			\tfrac{1}{2} D^2\,X^{;\alpha }X_{;\alpha }
		)
	]^\frac{1}{2} \sqrt{-g}\,\dd^4x.
	\nonumber
\end{align}
Because the coefficient of $ \sqrt{-g}\,\dd^4x $ on the right hand are not functional of $ (\phi ,X) $ alone, employing integration by parts in an attempt to remove some derivative terms or in dealing with residual derivative terms in the construction of Riemann/Ricci tensor, is unwieldy. 

In an attempt to make progress, let us suppose that we can perform series expansions for $ \widehat X $ and $ \sqrt{-\widehat g}  $ and decompose them as 
\begin{align}
	\label{decomxg}
	\widehat X
	&=
	\widehat X_s
	+
	\sum_{n = 1}^\infty \Big[
		c_n(\phi ^{;\alpha }X_{;\alpha })^n
		+ 
		d_n(X ^{;\alpha }X_{;\alpha })^n
	\Big],
	\nonumber
	\\[0.5em]
	\sqrt{-\widehat g} 
	&=
	\sqrt{-\widehat g_s}
	+
	\sum_{n = 1}^\infty \Big[
		p_n (\phi ^{;\alpha }X_{;\alpha })^n
		+ 
		q_n(X ^{;\alpha }X_{;\alpha })^n
	\Big],
\end{align}
where $ c_n, d_n, p_n, $ and $ q_n $ are functionals of $ (\phi ,X) $ only, and
\begin{align}
	\widehat X_s
	&=
	\frac{X}{A - 2BX},
	\nonumber
	\\[0.5em]
	\sqrt{-\widehat g_s}  
	&=
	A^\frac{3}{2}(A - 2BX)^\frac{1}{2} \sqrt{-g}.
\end{align}
The two terms $ \widehat X_s $ and $ \widehat g_s $ are the special disformal limit when $ \Phi _\mu = C\phi _{;\mu } + D X_{;\mu }\rightarrow  \sqrt{B}\phi _{;\mu } $.
The decomposition (\ref{decomxg}) rests on the assumption that the summations involving $ \phi ^{;\alpha }X_{;\alpha } $ and $ X^{;\alpha }X_{;\alpha } $ are small corrections $ \widehat X_s $ and $\sqrt{-\widehat g_s}  $. With this decomposition we may also decompose the transform functionals $ \widehat P, \widehat G_3, \widehat G_4 $ and $ \widehat G_5 $, and the derivatives $ G_{4,X} $ and $ G_{5,X} $, into their respective special disformal limit plus terms involving $ \phi , X, \phi ^{;\alpha }X_{;\alpha }, X^{;\alpha }X_{;\alpha }$, and other higher-order derivative terms. We may then express the transformed Horndeski action as a sum of actions of Horndeski form $ \mathcal S_{\mathcal H} $, and another one $ \mathcal S_{\mathcal B} $ (beyond Horndeski), involving derivative terms higher than two. The latter has the general dependencies given by 
\begin{align}
	\mathcal S _{\mathcal B}
	&=
	\mathcal S _{\mathcal B}\{
		\phi , X, \phi^{;\alpha }X_{;\alpha }, X^{;\alpha }X_{;\alpha }, R, R_{\mu \nu }, R^\alpha _{\mu \beta \nu },
		\nonumber
		\\[0.5em]
		&\qquad\quad 
		\text{higher order derivative terms}
	\}.
\end{align}

Following (\ref{decomxg}), $ \mathcal S_{\mathcal B} $ contains powers of $ \phi ^{;\alpha }X_{;\alpha }$ and $ X^{;\alpha }X_{;\alpha }$. The problematic terms include (but not limited to)
\begin{align}
	f(\phi ,X)(\phi^{;\alpha }X_{;\alpha })^n
	\quad \text{and}\quad 
	h(\phi ,X)(X^{;\alpha }X_{;\alpha })^n 
\end{align}
where $ f $ and $ h $ are coefficient functions in the expansion in terms of powers of $ \phi ^{;\alpha }X_{;\alpha }$ and $ X^{;\alpha }X_{;\alpha } $. Computing the Euler-Lagrange equations for these two, we find
\begin{align}
	&\left(
		\frac{\partial }{\partial \phi }
		-
		\nabla _\mu 
		\frac{\partial }{\partial \phi _{;\mu }}  
		+
		\nabla _\mu \nabla _\nu 
		\frac{\partial }{\partial \phi _{;\nu \mu }}
	\right) [f(\phi^{;\alpha }X_{;\alpha })^n]
	\nonumber
	\\[0.5em]
	&\quad=
	\cdots - n(n-1)f(\phi ^{;\alpha }X_{;\alpha })^{n-2}
	\phi ^{;\beta }\phi ^{;\mu }\phi ^{;\nu }
	X_{;\beta \mu \nu },
	\nonumber
	\\[0.5em]
	&\left(
		\frac{\partial }{\partial \phi }
		-
		\nabla _\mu 
		\frac{\partial }{\partial \phi _{;\mu }}  
		+
		\nabla _\mu \nabla _\nu 
		\frac{\partial }{\partial \phi _{;\nu \mu }}
	\right) [h(\phi^{;\alpha }X_{;\alpha })^n]
	\nonumber
	\\[0.5em]
	&\quad=
	\cdots - 4n(n-1)h(X^{;\alpha }X_{;\alpha })^{n-2}
	X^{;\beta }X^{;\mu }\phi ^{;\nu }X_{;\beta \nu \mu }
	\nonumber
	\\[0.5em]
	&\qquad\qquad\quad\; 
	-\,
	2nh(X^{;\alpha }X_{;\alpha })^{n-1}\phi ^{;\mu }
	\square X_{;\mu },
\end{align}
where we have only written the highest order derivative terms. In this case, for the two equations above, the order of derivative goes up to four for $ \phi  $. And generalizing $ A $ and $ B $ in (\ref{limgenextDT}) to have an additional dependence on $ (Y,Z) $ will only make matters worse both for the equation of motion and the transformation of the Horndeski action. In Ref. \cite{BenAchour:2016fzp}, the authors presented all degenerate higher order scalar tensor theories beyond Horndeski up to cubic powers of  second-order derivative of $ \phi $ in the action. Following the expansion given by (\ref{decomxg}) and its consequence on the transformed Horndeski action, we find upon performing integration by parts powers of second-order derivative of $ \phi $ higher than three. Moreover, degeneracy analysis following the logic in Ref. \cite{Langlois:2015cwa} is indicative that the transformed Horndeski sub-Lagrangians are non-degenerate. At the moment there is no easy solution in sight for the resulting Ostrogradsky instability and ghost(s) in the field quantization. We leave it for future work to look into the possibility of finding constraints to remove undesirable modes in the transformed Horndeski action. In addition to this, there may be some special forms of the functionals $ P, G_3, G_4, $ and $ G_5 $ so that the transformed Horndeski action is degenerate in nature. 
\end{multicols}

\section{Equations for $\widehat X$, $ \widehat Y $, and $ \widehat Z $}
\label{appendB}
In this appendix, we list down the full equations for the hatted quantities $(\widehat X,\widehat Y ,\widehat Z) $ in terms $ (\phi, X, Y, Z) $ and derivatives thereof.

By definition, $ X \equiv -\frac{1}{2} g^{\mu \nu }\phi _{;\mu }\phi _{;\nu } $. Its transformation under the general extended disformal transformation is given by 
\begin{align}
	\widehat X
	&=
	\frac{{{D}^{2}} \left( 2 X Z+{{Y}^{2}}\right) +2 A X}
	{2 A\, \left( {{D}^{2}} Z+2 C D Y-2 {{C}^{2}} X+A\right) }.
\end{align}

By definition, $ Y \equiv g^{\mu \nu }\phi _{;\mu }X_{;\nu } $. Its transformation under the general extended disformal transformation is given by
\begin{align}
	\widehat Y
	&=
	(AH_1)^{-1}\big[
		-H_{11} {H_7} {X}^{;\nu } Z_{;\nu }
		+
		H_{11} {H_6} {\phi}^{;\nu } Z_{;\nu }
		-
		H_{10} {H_7} {X}^{;\nu } Y_{;\nu }
		-
		H_{{{10}}} {H_6} {Z_2}	
		\nonumber
		\\
		&\qquad\quad
		-\,
		{\left( {H_7} {H_9}+2 {H_{{{10}}}} {H_6}\right)  Z}
		+
		{\left( {H_6} {H_9}-{H_7} {H_8}\right)  Y}
		-
		{2 {H_6} {H_8} X}
	\big],
\end{align}
where
\begin{align}
	{H_1}
	&=
	{{D}^{2}} Z+2 C D Y-2 {{C}^{2}} X+A,
	\nonumber
	\\[0.5em]
	{H_2}
	&=
	{{D}^{2}} \left( 2 X Z+{{Y}^{2}}\right) +2 A\, X,
	\nonumber
	\\[0.5em]
	{H_3}
	&=
	2 D\, \left( 2 X Z+{{Y}^{2}}\right),
	\nonumber
	\\[0.5em]
	{H_4}
	&=
	2 \left( D Z+C Y\right),
	\nonumber
	\\[0.5em]
	{H_5}
	&=
	2 \left( D Y-2 C X\right) ,
	\nonumber
	\\[0.5em]
	{H_6}
	&=
	{{D}^{2}} Z+C D Y+A,
	\nonumber
	\\[0.5em]
	{H_7}
	&=
	D\, \left( D Y-2 C X\right),
	\nonumber
	\\[0.5em]
	{H_8}
	&=
	\frac{{A_\phi} \left( 2 A {H_1} X-{H_1} {H_2}-A {H_2}\right) -A {C_\phi} {H_2} {H_5}+{D_\phi} \left( A {H_1} {H_3}-A {H_2} {H_4}\right) }{2 {{A}^{2}} {{{H_1}}^{2}}},
	\nonumber
	\\[0.5em]
	{H_9}
	&=
	\frac{1}{2 {{A}^{2}} {{{H_1}}^{2}}} \big[
		2 A\, \left( {{D}^{2}} {H_1} Z + {{C}^{2}} {H_2}+A {H_1}\right) 
		+
		{A_X} \left( 2 A {H_1} X-{H_1} {H_2}-A {H_2}\right)
		\nonumber
		\\
		&\qquad\qquad\qquad		
		-\,
		A {C_X} {H_2} {H_5}
		-
		A {D_X} \left( {H_2} {H_4}-{H_1} {H_3}\right) 
	\big],
	\nonumber
	\\[0.5em]
	H_{10}
	&=
	\frac{1}{2 {{A}^{2}} {{{H_1}}^{2}}} \big[
		2 A D\, \left( D {H_1} Y-C {H_2}\right) 
		+
		{A_Y} \left( 2 A {H_1} X-{H_1} {H_2}-A {H_2}\right)
		\nonumber
		\\
		&\qquad\qquad\qquad			
		-\,
		A {C_Y} {H_2} {H_5}
		-
		A {D_Y} \left( {H_2} {H_4}-{H_1} {H_3}\right) 
	\big],
	\nonumber
	\\[0.5em]
	H_{11}
	&=
	\frac{1}{2 {{A}^{2}} {{{H_1}}^{2}}}\big[
		{A_Z} \left( 2 A {H_1} X-{H_1} {H_2}-A {H_2}\right) 
		+
		A {{D}^{2}} \left( 2 {H_1} X-{H_2}\right)
		\nonumber
		\\
		&\qquad\qquad\qquad		
		-\,
		A {C_Z} {H_2} {H_5}
		-
		A {D_Z} \left( {H_2} {H_4}-{H_1} {H_3}\right) 
	\big].
\end{align}
The subscript of $ (A, B, C) $ denotes derivative; e.g., $ A_Y = \partial A/\partial Y $.

By definition, $ Z \equiv g^{\mu \nu }X_{;\mu }X_{;\nu } $. Its transformation under the general extended disformal transformation is given by
\begin{align}
	\widehat Z
	&=
	\frac{1}{AH_1} \big[
		-D^{2} {H_{11}}^2 X^{;\mu}  Z_{;\mu}  X^{;\nu}  Z_{;\nu} 
		-
		C D H_{11}^2 \phi^{;\mu}  Z_{;\mu}  X^{;\nu}  Z_{;\nu} 
		-
		D^2 H_{10} H_{11} X^{;\mu}  Y_{;\mu}  X^{;\nu}  {Z_{;\nu} }
		\nonumber
		\\
		&\qquad\quad
		-\,
		C D H_{11}^2X^{;\mu}  Z_{;\mu}  {{\phi}^{;\nu} } {Z_{;\nu} }
		-
		C^2 H_{11}^2\phi^{;\mu}  Z_{;\mu}  {{\phi}^{;\nu} } {Z_{;\nu} }
		-
		C D H_{10} H_{11} X^{;\mu}  Y_{;\mu}  {{\phi}^{;\nu} } {Z_{;\nu} }
		\nonumber
		\\[0.25em]
		&\qquad\quad		
		-\,
		D^2 H_{10} H_{11} X^{;\mu}  Z_{;\mu}  X^{;\nu}  {Y_{;\nu} }
		-
		C D H_{10} H_{11} \phi^{;\mu}  Z_{;\mu}  X^{;\nu}  {Y_{;\nu} }
		-
		D^2 {H_{10}^{2}} X^{;\mu}  Y_{;\mu}  X^{;\nu}  {Y_{;\nu} }
		\nonumber
		\\[0.25em]
		&\qquad\quad	
		+\,
		H_1H_{11}^2Z^{;\mu}  Z_{;\mu} 
		+
		2 H_1 H_{10} H_{11} Y^{;\mu}  Z_{;\mu} 
		+
		H_{12} X^{;\mu}  Z_{;\mu}
		+
		H_{13}  \phi^{;\mu}  Z_{;\mu}	
		\\[0.5em]
		&\qquad\quad	 
		+\,
		H_1{H_{10}^{2}} Y^{;\mu}  Y_{;\mu}
		+
		H_{14} X^{;\mu}  Y_{;\mu} 		
		+
		H_{15} \left( Z_2+2 Z\right)
		-
		H_1H_{10} H_8 \left( Z_2+2 Z\right) 
		\nonumber
		\\[0.5em]
		&\qquad\quad		
		+\,
		C D H_{10} H_9 Z Z_2
		+
		C^2 H_{10} H_9 Y Z_2
		+
		C D H_{10} H_8 Y Z_2
		-
		2 C^2 H_{10} H_8 X Z_2
		\nonumber
		\\[0.5em]
		&\qquad\quad		
		-\,
		H_9 Z\, \left( D^2 H_9 Z-2 C D H_{10} Z+C D H_9 Y+D^2 H_8 Y-2 C D H_8 X-H_1 H_9\right) 
		\nonumber
		\\[0.5em]
		&\qquad\quad		
		-\,
		H_8 Y\, \left( D^2 H_9 Z-2 C D H_{10} Z+C D H_9 Y+D^2 H_8 Y-2 C D H_8 X-H_1 H_9\right) 
		\nonumber
		\\[0.5em]
		&\qquad\quad		
		-\,
		H_9 Y\, \left( C D H_9 Z-2 C^2 H_{10} Z+C^2 H_9 Y+C D H_8 Y-2 C^2 H_8 X-H_1 H_8\right) 
		\nonumber
		\\[0.5em]
		&\qquad\quad		
		+\,
		2 H_8 X\, \left( C D H_9 Z-2 C^2 H_{10} Z+C^2 H_9 Y+C D H_8 Y-2 C^2 H_8 X-H_1 H_8\right) 
		\nonumber
	],
\end{align}
where
\begin{align}
	H_{12}
	&=
	2 H_{11} \left( C D H_{10} Z_2-D^2 H_9 Z
		+
		2 C D H_{10} Z-C D H_9 Y-D^2 H_8 Y\right.
		\\[0.25em]
		&\qquad\qquad		
		\left.
		+\,
		2 C D H_8 X+H_1 H_9
	\right),
	\nonumber
	\\[0.5em]
	H_{13}
	&=
	2 H_{11} \left( C^2 H_{10} Z_2-C D H_9 Z
	+
	2 C^2 H_{10} Z-C^2 H_9 Y-C D H_8 Y\right.
		\nonumber
		\\[0.25em]
		&\qquad\qquad		
		\left.		
		+\,
		2 C^2 H_8 X+H_1 H_8
	\right)	,
	\nonumber
	\\[0.5em]
	H_{14}
	&=
	2 H_{10} \left( C D H_{10} Z_2 
		- 
		D^2 H_9 Z+2 C D H_{10} Z-C D H_9 Y-D^2 H_8 Y
		\right.
		\nonumber
		\\[0.25em]
		&\qquad\qquad		
		\left.					
		+\,
		2 C D H_8 X+H_1 H_9
	\right),
	\nonumber
	\\[0.5em]
	H_{15}
	&=
	H_{10} \left( C D H_9 Z-2 C^2 H_{10} Z+C^2 H_9 Y+C D H_8 Y-2 C^2 H_8 X-H_1 H_8\right).
	\nonumber	 		
\end{align}

\begin{multicols}{2}

\end{multicols}
\end{document}